\documentclass[final,11pt]{elsarticle}
\usepackage{amssymb}
\usepackage{hyperref}
\usepackage[a4paper, total={6.6in, 9.8in}]{geometry}
\usepackage[utf8]{inputenc}
\usepackage[T1]{fontenc}
\usepackage{cite}
\usepackage{amsmath,amssymb,amsfonts}
\usepackage{algorithmic}
\usepackage{algorithm}              
\usepackage{graphicx}
\usepackage{textcomp}
\usepackage{subfigure}
\usepackage{xcolor}
\usepackage{soul}
\usepackage{url}
\usepackage{epstopdf}
\usepackage{multirow}
\def\BibTeX{{\rm B\kern-.05em{\sc i\kern-.025em b}\kern-.08em
    T\kern-.1667em\lower.7ex\hbox{E}\kern-.125emX}}

\usepackage{tikz}
\usepackage{tikz-3dplot} 
\definecolor{gold}{rgb}{0.85,.66,0}
\definecolor{cian}{rgb}{.02,.7,.95}
\definecolor{ppp}{rgb}{.7,.3,.82}

\tdplotsetmaincoords{60}{125}
\tdplotsetrotatedcoords{0}{0}{0} 
\DeclareMathOperator{\acos}{acos}

\journal{Arxiv (\href{https://arxiv.org/}{https://arxiv.org/})}

\begin{document}

\begin{frontmatter}
\title{\textbf{\LARGE GA-Aided  Directivity in Volumetric and Planar Massive-Antenna Array Design}}
\author{\textbf{Bruno Felipe Costa, Taufik Abrão}\\
{Londrina State University (UEL)},\\
{Department of Electrical Engineering}, 
{Londrina},
{86057-970}, 
{Parana},
{Brazil}}


\begin{abstract} 
{The problem of directivity enhancement, leading to the increase in the directivity gain over a certain desired angle of arrival/departure (AoA/AoD), is considered in this work.} A new formulation of the volumetric array directivity problem is proposed using the rectangular coordinates to describe each antenna element, and the desired {azimuth and elevation angles with a general element pattern. Such a directivity problem is formulated to find the optimal minimum distance between the antenna elements $d_\text{min}$ aiming to achieve as high directivity gains as possible.}  {An expedited implementation method is developed to place the antenna elements in a distinctive plane dependent on ($\theta_0$; $\phi_0$).}  A novel concept on optimizing directivity for the uniform planar array (OUPA) is introduced to find a quasi-optimal solution for the non-convex optimization problem with low complexity.  This solution is reached by deploying the proposed successive evaluation and validation (SEV) method.  {Moreover, the genetic} algorithm (GA) method was deployed to find the directivity optimization solution expeditiously. For a small number of antenna elements {, typically $N\in [4,\dots, 9]$,} the achievable directivity by GA optimization demonstrates gains of $\sim 3$ dBi compared with the traditional beamforming technique, using steering vector for uniform linear arrays (ULA) and uniform circular arrays (UCA), while gains of  $\sim1.5$ dBi is attained when compared with an improved UCA directivity method. For a larger number of antenna elements {, two improved GA procedures, namely GA-{\it marginal} and GA-{\it stall}, were} proposed and compared with the OUPA method. OUPA also indicates promising directivity gains surpassing $30$ dBi for massive MIMO scenarios. 
\end{abstract}
\end{frontmatter}

%
\section{Introduction}  
Array antennas can provide many advantages in flexibility and scalability concerning conventional antennas. This is related to the capacity of array pattern reconfiguration and adaptability.  Such adaptability enables various applications, including radar communications, satellite communications, wireless communications, radio-astronomy, remote sensing, and direction of arrival (DoA) estimation. Optimally designed arrays are used to obtain high directivity patterns, improving the array performance. 

Uniform array configurations, such as uniform linear (ULA), planar (UPA), rectangular (URA), circular or cylindrical  (UCA) antenna arrays, have been deployed to improve the performance of massive antenna-based 5G and B5G communication systems.  Since planar antenna array can deploy a higher number of antenna elements at a limited physical space and capture the beamspace in both the horizontal and vertical directions in three-dimensional (3D) propagation, the development of different planar array configurations, such as  URA and UCA, have become a natural choice among the different possible array geometries, representing a great interest in the mmWave massive MIMO system designs.

Array directivity optimization remains an open issue,  especially in cases where the antenna-element position is considered as an optimization variable. Indeed, the choice of antenna-elements position as a parameter to be optimized can be justified by the concept of {\it array virtualization}, in which antenna-elements cooperate among each other, forming a {\it virtual antenna array} (VAA) with better directivity performance that the original physical antenna array.

Recent works on array optimization have designed {\it linear sparse array} antennas. For instance, the algorithm proposed in \citep{Angeletti_2014a} jointly and deterministically maximizes the aperture efficiency and directivity. Besides, \citep{Angeletti_2014b} proposes a planar circular sparse array antenna design, where the positions and dimensions of the radiators are jointly optimized. In \citep{Bouchekara_2018} an optimal circular antenna array for maximum sidelobe levels (SLLs) reduction was investigated using a metaheuristic approach called {\it player algorithm}. An optimization of sidelobe level and aperture efficiency for aperiodic array antennas was proposed in \citep{Diao_2017}. Multiobjective optimization (MOO) is deployed in \citep{Goudos_2013} to efficiently and effectively design subarrays in linear antenna array; the optimization was made using a {\it memetic differential evolution} (mDE)  algorithm. Following the same research direction, the authors in \citep{Jayaprakasam_2014, Jayaprakasam_2017} used MOO and metaheuristic methods to optimize the beampattern in collaborative array beamforming design. In \citep{Tripathy_2019}, the MOO technique is deployed to discuss the planar antenna array design under different optimization criteria, such as the side lobe, beam steering quality, minimum power beam width, and maximum directivity, as a function of element lengths and distances between these elements. A fundamental question is exploring the Pareto front modeling optimal trade-offs among multiple conflicting objectives, including the structure size, its performance, and the minimization of cost.  The well-known heuristic algorithm {\it nondominated sorting genetic algorithm} II (NSGA-II) has been used for this purpose. Many other works are deploying MOO and/or metaheuristics solutions, such as \citep{Reddy_2018}, \citep{Rezagholi_2016}, \citep{Saleem_2016}, \citep{Swain_2016} and \citep{Yang_2015}. 

In \citep{Kumar_2017}, the optimal radiation pattern is generated using a {\it spherical phased array}. In contrast, in \citep{Huang_2016}, a technique based on {\it changes in the subspace} is deployed to improve the directivity of the omnidirectional UCA. Moreover,  the {\it optimum directive beamformer} has been proposed in \citep{Trucco_2014}, by deploying the concept of {\it generalized directivity maximization}; the concept of directivity is based on the {\it mean beam power pattern},  which is crucial in assessing and optimizing the performance of directivity arrays subjected to array imperfections and sensor mismatch. Recently, by adopting a 3D wireless channel model, and taking into account both the azimuth and elevation angles of departure (AoD), the authors of \citep{Tan2022} derive the probability density function for the distances between randomly distributed users and the BS antenna array, considering three URA, ULA and UCA uniform array topologies. Also, they derived closed-form expressions for the squared inner product of different channel vectors, facilitating the interuser interference analysis.

Against this background, the current work proposes a new methodology to maximize the directivity of omnidirectional volumetric antenna arrays.  By assuming a uniform planar array (UPA) confined on a specific plane with a minimum distance between the antenna elements optimized, the proposed method takes in hand the successive evaluation and validation (SEV) procedure. To demonstrate effectiveness, the proposed method is implemented by deploying evolutionary heuristic optimization;  hence, to implement the directivity optimization methodology under a large number of antennas in planar arrays, we suggested two variants of the genetic algorithm: GA-{\it marginal} and GA-{\it stall}, both made the GA a promising optimization tool to solve UPA directivity problem in such large scale antenna scenario. An exciting finding of the proposed methodology is that a plane space constraints the antenna-element positioning solutions as a function of the desired elevation and azimuth angles.

\vspace{2mm}
\noindent The \textbf{\textit{contribution}} of this work is fourfold:
\begin{itemize}
\item[{\bf a})]  formulation of optimization problem for directivity maximization;  the novelty of this design is brought by the optimization of each antenna-element position bounded by a volumetric constraint on the search space, which is given by a parallelepiped with a central point at the origin and all the points bounded;  
 \item[{\bf b})] a geometric interpretation for the solution in case of omnidirectional scenarios, more specifically, this interpretation results in restraining the position of each antenna element to a plane equation dependent on the desired angles ($\theta_0,\phi_0$); 
\item[{\bf c})] using a novel concept based on the UPA geometry, an expedited implementation method is developed consisting in placing the antenna elements in a distinctive plane dependent on $\theta_0$ and $\phi_0$; hence, using the geometry properties of the UPA, a directivity optimization problem was formulated to find the optimal minimum distance between the antenna-elements $d_\text{min}$ aiming to achieve as high directivity gains as possible;
 \item[{\bf d})] to corroborate the effectiveness of the proposed method, a genetic algorithm (GA) is deployed to implement the directivity enhancement method with low complexity; when compared with a conventional geometric design, the proposed method has demonstrated a significant improvement in directivity gains on the desired angle of departure (DoD).
\end{itemize}

The remainder of the paper is organized as follows.  Section \ref{sec:dire_opt}  describes the concept of the radiation pattern, directivity optimization via position antenna-element placement, and the adopted system model.  Section \ref{sec:opt} formulates the directivity optimization problem,  with particular attention on the optimization problem for the omnidirectional case; besides,  geometric interpretation is evoked, and then the problem is recast in a simplified way.  For comparison purposes in terms of directivity, different geometric array structures are considered. 
A novel directivity optimization method applied to the UPA geometries is devised in Section \ref{sec:algorithm} as an effective low-complexity optimization method. Numerical results are developed in Section \ref{sec:numerical} by deploying heuristic GA as a practical implementation tool supporting our finding on the UPA-basis antenna array directivity optimization technique. Concluding remarks are pointed out in Section \ref{sec:concl}.

\section{Radiation Pattern and Directivity via Element Positioning}\label{sec:dire_opt}
\begin{figure}[!htbp]
 \centering
 \includegraphics[width=.9\textwidth]{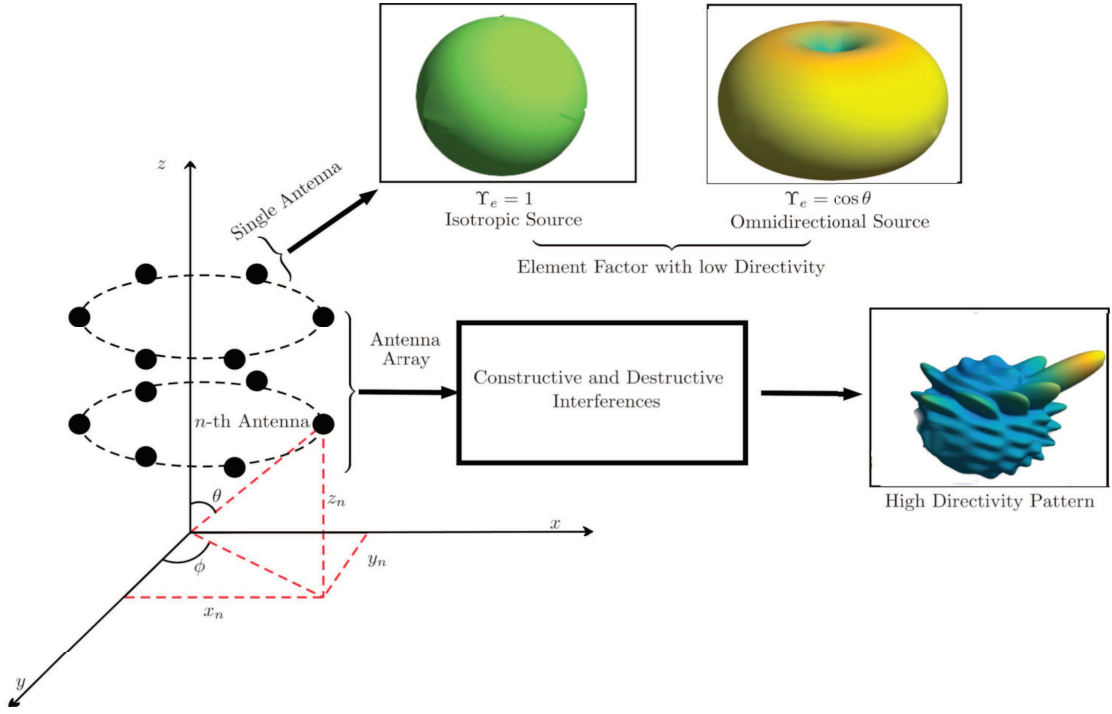}
 \vspace{-4mm}
 \caption{Directivity representation in a) isotropic source, b) omnidirectional source, and c) via constructive and destructive interferences of low-directivity sources.}
 \label{fig:02}
 \end{figure}

For non-isotropic antennas, the directivity is the level of radiated signal power in a specific direction. This is especially important for applications such as beamforming in massive MIMO 5G wireless communications. Fig. \ref{fig:02} depicts the general concept of directivity enhancement via an array of antenna elements with low directivity; the constructive and destructive interference can be guided, by changing the position and/or phase of each element, aiming to achieve a high directivity antenna pattern \citep{BFC_2018}. In this work, the omnidirectional antenna-element configuration has been chosen to implement the {array} directivity {enhancement}, which is a more realistic scenario compared to the isotropic sources, while remaining a {suitable} configuration to be {analytically} analyzed due to its simple element factor, expressed by $\Upsilon(\theta) = \cos{\theta}$. Furthermore, the applicability of this configuration is interesting in many fields.  Such configuration dramatically simplifies the analytical directivity expression, while the non-convex directivity optimization problem remains a special issue.

\subsection{Radiation Pattern of an Antenna Array}\label{sec:pattern}
An antenna array is a grouping of $N$ antennas whose geometry can assume many different forms, such as uniform linear, circular, rectangular, non-uniform grouping, and so forth; moreover, all elements in the array work jointly to increase the directivity of the arrangement.  The entire field of an antenna array is obtained by multiplying the field of a single element $\Upsilon_e (\theta, \phi)$ and the array factor $\Upsilon_a (\theta, \phi)$; this is called the radiation pattern of an antenna array and can be written as:
\begin{equation}
\Upsilon (\theta , \phi)  = \Upsilon_e (\theta , \phi) \Upsilon_a (\theta , \phi)
\label{eq:1}
\end{equation} 
The element factor (EF) is the radiation pattern of a single-element antenna, and the model must resemble the pattern of a real antenna. Using spherical coordinates, the EF is periodic in $\theta$ and $\phi$ angles. Thus, the radiated power of an arbitrary antenna can be described using Fourier analysis, {\it i.e.} it can be written using a linear combination of powers of cosine and sine functions. Hence, the  EF can be written:
\begin{equation}
\Upsilon_e (\theta)  = \sin^u{(\theta)}\cos^v{(\theta)}
\label{EF}
\end{equation}

The {\it array factor} (AF) describes a combination of radiating elements in an array without considering the element radiation pattern. Thus, the array factor can be interpreted as the radiation pattern by replacing the actual elements by the isotropic (point) sources, \textit{i.e.},  $\Upsilon_e(\theta,\phi) = 1$. The AF has a dependency in terms of position, relative phase and relative amplitude of each antenna element and is given by:
\begin{equation}
\Upsilon_a (\theta,\phi) = \sum \limits_{n=1}^N A_ne^{j(\alpha_n + k \boldsymbol{p}_n \cdot \boldsymbol{a}_p)}
\label{arraygeral}
\end{equation}
where $N$ is the number of antenna elements, $A_n$ is the relative amplitude of $n$-th antenna element, $\alpha_n$ is the relative phase of $n$-th antenna element, $\boldsymbol{p}_n$ is the position vector of $n$-th antenna element; $k$ is the wave number, and $\boldsymbol{a}_p$ is the unit vector of observation point in spherical coordinates, given by:
\begin{equation}
\boldsymbol{a}_p = \sin (\theta) \cos(\phi) \hat{i} + \sin (\theta) \sin(\phi)  \hat{j} + \cos (\theta) \hat{k}
\label{ar}
\end{equation}

The position vector can be written as a combination of the elements in the 3-D cartesian axis:
\begin{equation}
\boldsymbol{p}_n = {x_{n}} \hat{i} + {y_{n}} \hat{j} + {z_{n}} \hat{k}
\label{position}
\end{equation}
In this work, we have considered the position of each antenna element of the array as the variable to be optimized, which the matrix can mathematically defined:
\begin{equation}\label{eq:P}
{\boldsymbol{P} = \left  [\begin{array}{c} \boldsymbol{p}_1\,\, \boldsymbol{p}_2 \,\, \hdots  \,\,\boldsymbol{p}_N\end{array} \right]^{T}}
\end{equation}
where the position vector $\boldsymbol{p}_n$ defines the localization of the $n$-th antenna element in the 3-D space.

\subsection{Directivity for Arbitrary Volumetric Antenna Arrays}\label{sec:expression}
Directivity can be defined as the level of irradiation signal power in a specific direction to the detriment of others. This is especially important for some applications, such as beamforming in non-isotropic massive MIMO antennas in 5G wireless communications.  Antenna directivity can be written as the ratio between the radiation intensity  in a desired angle and the sum of the radiation intensity in all the other directions:
\begin{equation}
\mathcal{D}(\theta_0 , \phi_0) = \frac{|\Upsilon (\theta_0 , \phi_0)|^2 }{\frac{1}{4\pi} \int \limits_{0}^{2\pi} \int \limits_{0}^{\pi} |\Upsilon(\theta , \phi) |^2\sin{(\theta)} d\theta d\phi}
\label{eq:direnorm} 
\end{equation}
where $|\Upsilon (\theta , \phi)|^2$ is the radiation intensity.   Besides, for an arbitrary array with the general element pattern ($\cos^v{\theta}\sin^u{\theta}, \,\, \forall \, \, u, v$), the directivity expression can be defined as 
\citep{BFC_2018}:
\begin{align}
\begin{split}
&\mathcal{D}(\theta_0 , \phi_0, u,v)= \frac{f_1}{f_2}=
 \frac{\sin^{2u}{(\theta_0)}\cos^{2v}{(\theta_0)} \sum\limits_{\substack{m,n=1}}^N A_n A_m \xi_{mn}\cos{\left[\Omega_{mn}\right]} }{\mathcal{I}_n(u,v)+ \mathcal{I}_m(u,v)  } 
 \end{split}
 \label{equation_geral}
 \end{align}
where
\begin{align}
\begin{split}
& \mathcal{I}_n(u,v)= \sum \limits_{n=1}^N A_n^2 \left[ \frac{1}{8}((-1)^{2v}+1)\mathcal{B}(u+1,v+\frac{1}{2}) \right] \\
& \mathcal{I}_m(u,v) = 2 \, (-1)^{(v+2u)} \sum\limits_{\substack{n,m=1 \\ m\neq n \\ n > m}}^N \sum \limits_{\kappa = 0}^u A_n A_m \binom{u}{\kappa} \cos{(\alpha_{mn})}\frac{\partial^{2(v+u-\kappa)}}{\partial z_{mn}^{2(v+u-\kappa)}} \left [\frac{\sin (k\sqrt{\beta^2+z_{mn}^2})}{k\sqrt{\beta^2+z_{mn}^2}} \right ] \\
&\Omega_{mn} =\Omega(\boldsymbol{p}_n,\boldsymbol{p}_m,\alpha_n,\alpha_m,\theta_0,\phi_0) = k \, [x_{nm}\sin{\theta_0} \cos{\phi_0} + y_{nm} \sin{\theta_0} \sin{\phi_0} + z_{nm} \cos{\theta_0} ] + \alpha_{nm} \\
&\boldsymbol{p_n} = x_n \hat{i} + y_n \hat{j} + z_n \hat{k} ,\hspace{0.5cm} \beta =  \sqrt{x_{nm}^2 + y_{nm}^2},\hspace{0.5cm} x_{nm} = (x_n-x_m), \hspace{0.5cm} y_{nm} = (y_n-y_m), \\
& \hspace{1cm} z_{mn} =({z_{n}}-{z_{m}}) \\
&\xi_{mn}  = \begin{cases} 1 \qquad \qquad \,\, m\neq n \\ \frac{1}{2 \cos{\left[\Omega_{mn}\right]}} \quad m=n \end{cases}, \hspace{0.5cm} \mathbb{R}e\{v\} > -\frac{1}{2},\hspace{0.5cm} {\mathbb{R}e}\{u\} > -1,\hspace{0.5cm} \alpha_{mn} =({\alpha_{n}}-{\alpha_{m}}). 
\label{eq_geral:def}
\end{split}
\end{align}

\vspace{2mm}
\noindent {where} $\boldsymbol{p_n}$ is the position vector of $n$-th antenna element;
$\alpha_n$ is the phase of $n$-th antenna element; $A_n$ is the  amplitude of $n$-th antenna element; $\theta_0$ is the desired elevation angle, where directivity is evaluated; $\phi_0$ represents the desired azimuth angle, where directivity is evaluated; and $k$ is the wave number of the transmission.

\section{Directivity Optimization Problem in Omnidirectional Scenarios} \label{sec:opt}
Given a desired angle of departure $(\theta_0 , \phi_0) $ over an antenna array composed of $N$ elements, we are looking for the parameter values of an array configuration that result in the directivity maximization in such an AoD.  

We start the analysis assuming non-zero-phase and the non-unitary amplitude in each antenna; hence, we will solve the proposed problem to obtain a geometric configuration for each desired angle. For typical applications, the desired direction changes regularly, which becomes a problem, given that changes in the positions of antennas are physically impossible. However, in this formulation, the phase element was not considered, leaving space to adopt virtualization techniques, which highly depend on the phase elements and will be the subject of our future research.
 
Based on the directivity expression provided in eq. \eqref{equation_geral}, we state the following optimization problem:
\begin{equation}
\begin{aligned}
& \underset{\boldsymbol{P}}{\text{maximize}} && \mathcal{D}(\theta_0 , \phi_0) = \frac{f_1}{f_2} \\
&\text{subject to} &&  \boldsymbol{p}_n  \preceq  \boldsymbol{p}_{\max} \quad  \text{for} \quad n = 1,2,3,\dots, N \\
\end{aligned}
\label{Problem:P}
\end{equation}
where the negative semi-definite  $\boldsymbol{p}_n  \preceq \boldsymbol{p}_{\max}$ defines the 3-D parallelepiped volume bounded by $[x_{\max}, y_{\max}, z_{\max}]$. Besides, the optimization problem in \eqref{equation_geral} can be rewritten as a minimization problem: 
\begin{equation}
\begin{aligned}
& \underset{\boldsymbol{P}}{\text{minimize}}
& & \frac{f_2(\boldsymbol{P}, \theta_0 , \phi_0)}{f_1(\boldsymbol{P}, \theta_0 , \phi_0)} \\
&\text{subject to} &&  \boldsymbol{p}_n  \preceq  \boldsymbol{p}_{\max} \quad  \text{for} \quad n = 1,2,3,\dots, N \\
\end{aligned}
\label{fracProblem}
\end{equation}
Therefore, the objective of this problem can be reinterpreted as minimizing $f_2(\boldsymbol{P})$. At the same time, simultaneously maximize $f_1(\boldsymbol{P})$ subject to the bounds in the space configuration,  given, for instance, by the 2-norm.

The  \textbf{\textit{omnidirectional scenario}} has been selected due to the concise expressions, analysis, and manageable complexity while the applicability remains in many exciting fields. The omnidirectional radiated power is also scattered symmetrically but not equally in all directions, creating a 3-D torus radiation pattern, which can be formulated as the expression $\Upsilon(\theta) = \cos{\theta}$ for all azimuth angles ($\phi$).  
Hence, considering the element pattern given in eq. \eqref{EF}, for the omnidirectional scenario,  we set $u=0$ and $v=1$. Such condition simplifies the auxiliary functions as:
\begin{align}\label{eq:f1-2}
f_{1}^{\textsc{o}}(\theta_0) =  \cos^{2}{(\theta_0)} \sum\limits_{\substack{m,n=1}}^N A_n A_m \xi_{mn}\cos{\left[\Omega_{mn}\right]}
\end{align}
and
\begin{align}\label{eq:f2-2}
f_2^{\textsc{o}} =  \sum \limits_{n=1}^N \frac{A_n^2}{6}  -2 \sum\limits_{\substack{m,n=1 \\ m\neq n\\ n > m}}^N A_n A_m \frac{\partial^{2}}{\partial z_{mn}^{2}} \left [\frac{\sin (k\sqrt{\beta^2+z_{mn}^2})}{k\sqrt{\beta^2+z_{mn}^2}} \right ] 
\end{align}
\begin{align}\label{dmn}
\text{where}\hspace{30mm} d_{mn} = k\sqrt{\beta^2+z_{mn}^2}
\end{align}
Generically, $d_{mn}$ represents  the Euclidean distance between the $m$-th and $n$-th antenna element in the 3-D space, given by $d_{mn} = k \sqrt{x_{mn}^2+y_{mn}^2+z_{mn}^2}$. \

In Eq. \eqref{Problem:P}, we are interested in maximizing $f_1^{\textsc{o}}$ while minimize $f_2^{\textsc{o}}$.  For this purpose, we can verify in Eq. \eqref{eq:f1-2}  that the only way to ensure these conditions consists in maximizing $\cos{\left[\Omega_{mn}\right]}$, {\it i.e.} 
\begin{align}
\Omega_{mn} = c_1\pi, \quad \text{for }\,\, \frac{c_1}{2} \in \mathbb{Z}\quad c_1\,\, \text{even}
\end{align}
Using the definitions in \eqref{eq_geral:def}, and satisfying the condition above, for $\Omega_{mn}$ we have:
\begin{align}
\begin{split}
&x_{nm} \sin{\theta_0} \cos{\phi_0} + y_{nm} \sin{\theta_0} \sin{\phi_0}
+ z_{nm} \cos{\theta_0} \,\, =\,\, \frac{c_1\pi - (\alpha_n -\alpha_m)}{k}
\end{split}\label{c1}
\end{align}
Eq. \eqref{c1} can be interpreted as the plane equation containing all the vectors of position difference ($\boldsymbol{p}_{mn} = \boldsymbol{p}_n - \boldsymbol{p}_m$), to fulfill this condition entirely it is necessary all vectors be restrained in the same plane of the differences position vectors since the constant $c_1$ is not unique. For each value, we have a different plane, the solution of this condition is given by infinite parallel planes. However, the combination of position vectors in these parallel planes does not fulfill the condition, given that the difference vector for points in different parallel planes will result in a vector difference in another plane. 

Moreover, minimizing $f_2^{\textsc{o}}$ results in a more complicated task, given that the directivity problem in \eqref{eq:direnorm} requires minimization of the summation of all terms in the right-hand side of \eqref{eq:f2-2}, instead of minimizing each term. Hence, the $f_2^{\textsc{o}}$ minimization problem can be recast in an alternative version as:
\begin{align} \label{c2}
\underset{\boldsymbol{P}}{\text{maximize}} \quad &\sum\limits_{\substack{m,n=1 \\ m\neq n \\ n > m }}^N A_n A_m \frac{\partial^{2}}{\partial z_{mn}^{2}} \left [\frac{\sin (k\sqrt{\beta^2+z_{mn}^2})}{k\sqrt{\beta^2+z_{mn}^2}} \right ]
\end{align}
Besides, the second derivative in \eqref{c2} can be re-written as:
\begin{align}\label{eq:betadist}
\begin{split}
\frac{\partial^{2}}{\partial z_{mn}^{2}} \left[\frac{\sin (\sqrt{\beta^2+z_{mn}^2})}{\sqrt{\beta^2+z_{mn}^2}} \right ] \,\, =\,\, &\,\, \frac{(\beta^2-2z_{mn}^2)\cos{d_{mn}}}{d_{mn}^4} \,\,\\ 
&\,\, -\frac{\left[(\beta^2-2)z_{mn}^2+\beta^2+z_{mn}^4\right]\sin{d_{mn}}}{d_{mn}^{5}}
\end{split}
\end{align}
which is function of the $x_{mn}$, $y_{mn}$ and $z_{mn}$, and related to $\beta$ and $d_{mn}$ through the  constraints in eq. \eqref{equation_geral} and \eqref{dmn}, respectively. Notice that in this formulation, the omnidirectional configuration has influence only in \eqref{eq:betadist}. For another scenarios, i.e., different values of $u,v\neq 0$, the only change will be the objective function (OF) in \eqref{c2}, where the order and number of derivatives dependent on the values  of $u, v\in\mathbb{Z}_+$.

\subsection{Recasting the Simplified Omnidirectional Problem}\label{sec:restraints}
One of the most important constraints resulting by the omnidirectional scenario is given in eq. \eqref{c1}, and where the values of $c_1$ can be chosen arbitrarily; hence, aiming to simplify the analysis we set $c_1 = 0$. Furthermore, considering all null phases, for the $n$-th and $m$-th antennas, we obtain:
\begin{align}\label{subs}
\begin{split}
&\sin{\theta_0}\cos{\phi_0}x_n + \sin{\theta_0}\sin{\phi_0}y_n + \cos{\theta_0}z_n = 0\\
&\sin{\theta_0}\cos{\phi_0}x_m + \sin{\theta_0}\sin{\phi_0}y_m + \cos{\theta_0}z_m = 0\\
\end{split}
\end{align}
By subtracting both previous expressions, one can obtain: 
\begin{align}\label{zmn}
z_{nm}= \tan{\theta_0}(\cos{\phi_0}x_{nm} + \sin{\phi_0}y_{nm})
\end{align}
The restriction \eqref{zmn} can be incorporated into \eqref{eq:betadist}, finally resulting in a simplified optimization problem:
\begin{equation}
\begin{aligned}
& \underset{\boldsymbol{x}, \boldsymbol{y}}{\text{minimize}}
& & \mathcal{G}(\boldsymbol{x},\boldsymbol{y}) = - \sum\limits_{\substack{m,n=1 \\ m\neq n \\ n > m}}^N \mathcal{F}(x_{mn},y_{mn} ) \\
&\text{subject to} &&  |x_{mn}| \,  \leq \, x_{\max}, \quad |y_{mn} |\,  \leq \, y_{\max} \quad  \text{for} \quad n = 1,2,3,\dots, N  \\
\end{aligned}
\label{newproblem}
\end{equation}
with OF defined by \eqref{eq:betadist}, and substituting $\beta^2 = \frac{d_{mn}^2}{k} - z_{mn}^2$, results:
\begin{align}
\begin{split}
\mathcal{F}(x_{mn},y_{mn}) \,\,=\,\, &\frac{\left[(d_{mn}^2-3k)z_{mn}^2+d_{mn}^2\right]\sin{d_{mn}}}{kd_{mn}^{5}}
 - \frac{(d_{mn}^2-3kz_{mn}^2)\cos{d_{mn}}}{kd_{mn}^4}
\end{split}\label{fdef}
\end{align}
The function $\mathcal{G}$ can be rewritten using the definition in \eqref{eq:f2-2}, implying the following relation:
\begin{align}
f_2^{\textsc{o}} =  \sum \limits_{n=1}^N \frac{A_n^2}{6}  +2 \mathcal{G}
\end{align}
Since $f_2^{\textsc{o}}$ represents the denominator integral in \eqref{eq:direnorm}, for the omnidirectional scenario this values must be positive; therefore, the function $\mathcal{G}$ has a lower bound for each value of $N$, given by:
\begin{align}\label{eq:Gbound}
\mathcal{G}_{\text{bound}}^N =  - \sum \limits_{n=1}^N \frac{A_n^2}{12}  
\end{align}
The bound on $z_{mn}$ is implicit, having dependency on $x_{\max}$, $y_{\max}$, and on the desired angle of departure $(\theta_0 , \phi_0) $. The solution of \eqref{newproblem} give us the optimal coordinates $x_n$ and $y_n, \,\, \forall n$  which can be used to find the optimal  $z_n, \,\, \forall n=1,\ldots, N$ coordinate  from \eqref{zmn}.
 
\subsection{Geometric Planar Arrays Comparison} \label{sec:Comparison}
In this section, different geometric planar array structures are considered for a directivity comparison purpose. It is shown the importance of setting the adequate $d_\text{min}$ and the associated area occupied by the array, aiming to exploit the potential of the directivity improvement. All the selected arrays are planar, aiming to preserve the constraint in \eqref{zmn}. Adopted parameter values and antenna structure include: 

\begin{center}
\small
\begin{tabular}{ll}
\hline
\bf Parameter & \bf Adopted Values \\
\hline
\# antennas  & $N=16$\\
Normalized wavelength & $\lambda=1$ [m] \\	
Desired direction & $\theta_0 = \phi_0 = \pi/4$\\
\hline
uniform planar array &UPA\\
uniform circular array & UCA\\
uniform hexagonal planar array & UHPA\\
\hline
\end{tabular}
\end{center}

\begin{figure*}[!htbp]
\centering
\subfigure[Uniform Planar Array (UPA)]{\includegraphics[width=.493\textwidth]{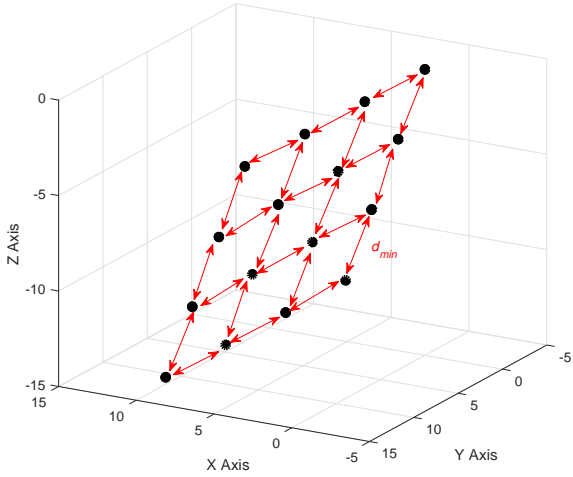}} 
\subfigure[Uniform Circular Array (UCA)]{\includegraphics[width=.493\textwidth]{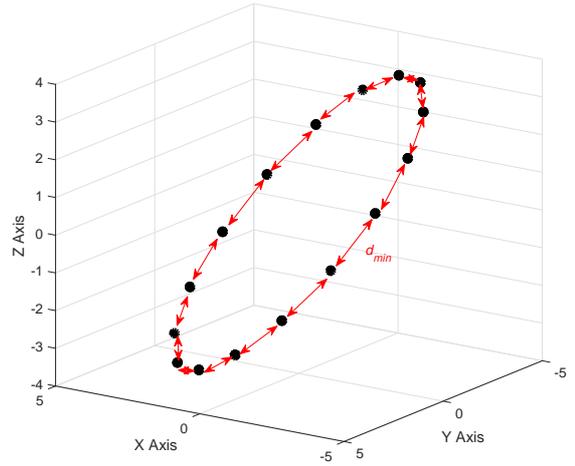}} 
\subfigure[Uniform Hexagonal Planar Array (UHPA)]{\includegraphics[width=.496\textwidth]{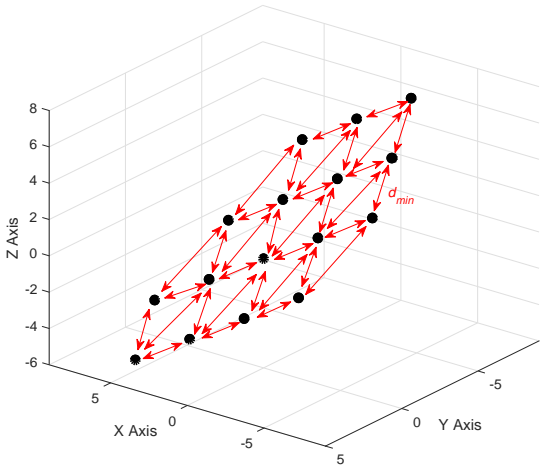}} 
\subfigure[Directivity {\it versus} the minimum distance for the three different geometry arrays: UPA, UCA, and UHPA]{\includegraphics[width=.463\textwidth]{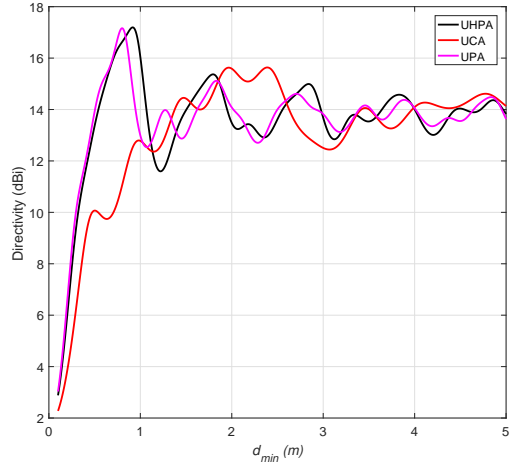}\label{fig:directivity}}
\caption{Different planar array geometries and respective attained directivity with $N=16$ elements on the desired plane defined by $\theta_0 = \phi_0 = \pi/4$.}
\label{fig:04}
\end{figure*}

The selected uniform planar array geometries with $N=16$ elements are depicted in Fig. \ref{fig:04} with the minimum distance between antenna elements $d_\text{min}$ represented by red arrows. Fig. \ref{fig:directivity} exhibits the directivity dependence regarding the increasing $d_\text{min}$ values for each selected uniform array geometries. 
The planar and the hexagon geometry attained superior directivity than the circular array; the optimal value of $d_\text{min}$ for both structures are close to each other and significantly smaller than the circular array {$d_\text{min}$ value}. For this application, the UCA has demonstrated do not be suitable and will be suppressed hereafter in further analyses. 

\begin{figure}[htbp!]
\centering
\subfigure[Directivity {\it vs.} the number of antennas increasing $(N)$]{\includegraphics[width=.495\textwidth]{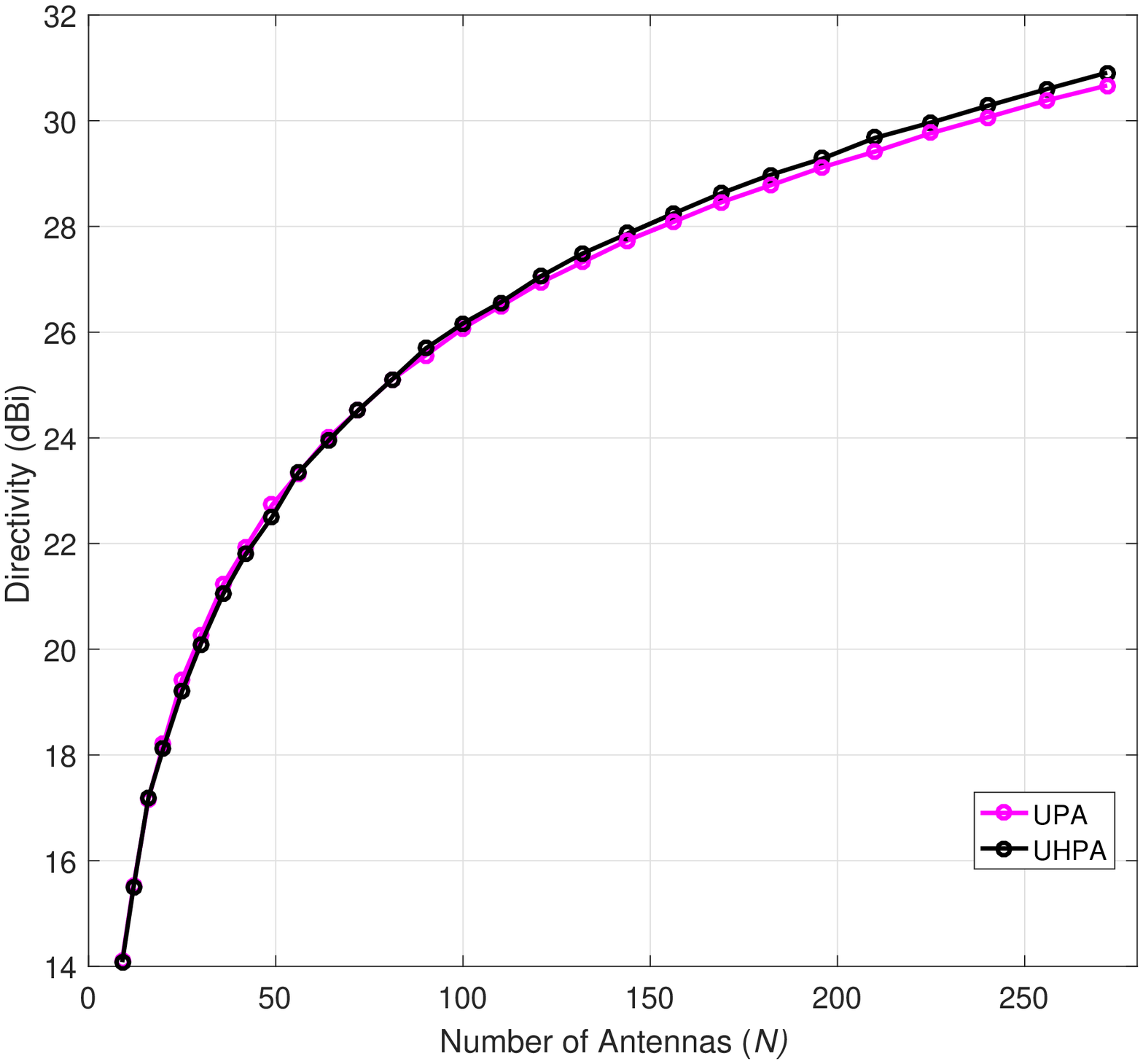}} 
\subfigure[Array area {\it vs.} the number of antennas $(N)$ increasing.]{\includegraphics[width=.495\textwidth]{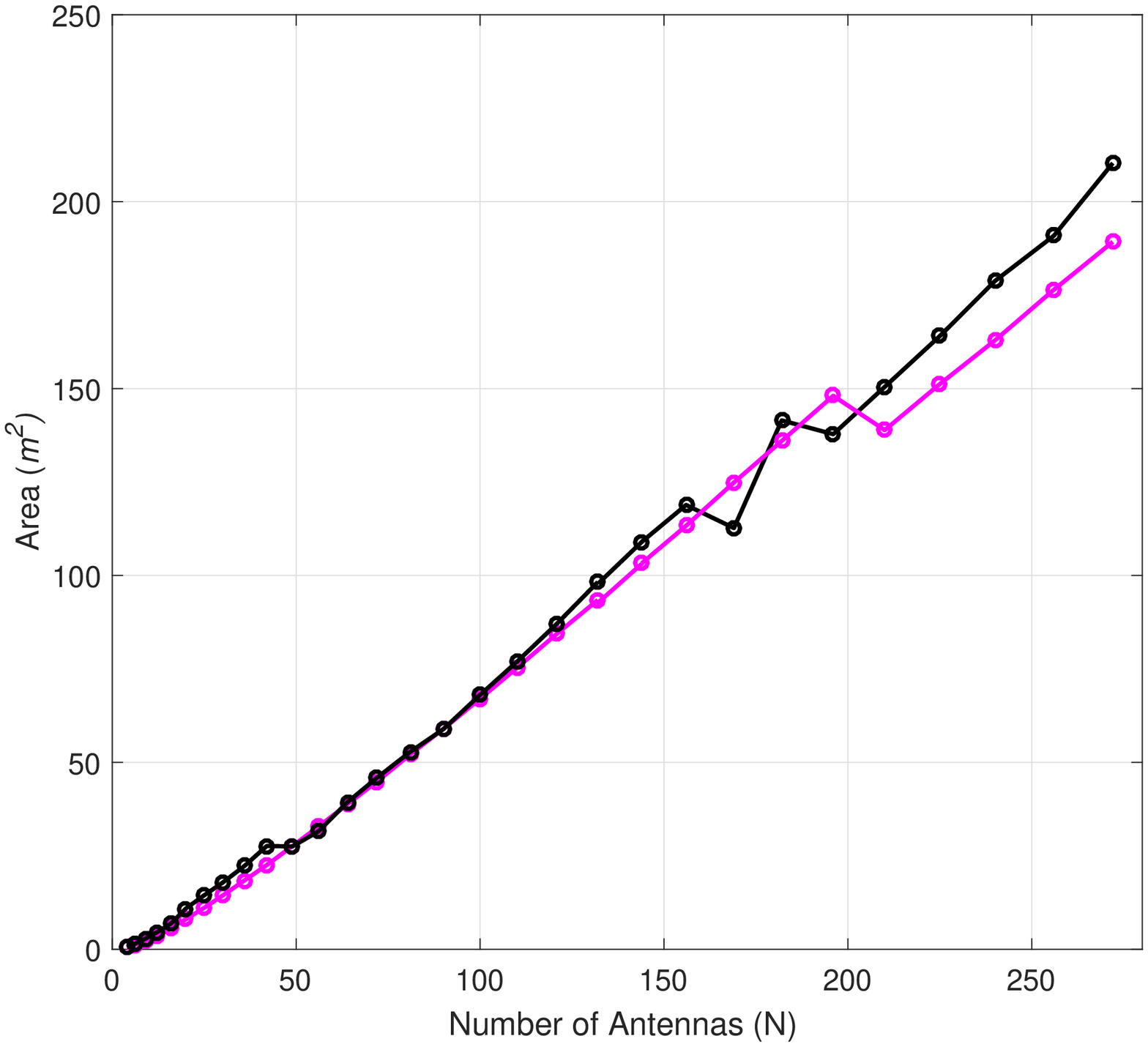}}
\vspace{-5mm}
\caption{UPA against UHPA in terms of directivity and area, both with optimal setting of $d_\text{min}$.}
\label{fig:06}
\end{figure}

These illustrative results express the importance of setting the adequate $d_\text{min}$ to exploit the directivity enhancement's potential. However, the area occupied by the array is also essential, since compact antenna arrays are more valuable. Although the value of $d_\text{min}$ is smaller for the UPA, the area is not necessarily smaller than the UHPA.  To investigate this aspect, Fig. \ref{fig:06} depicts directivity and the normalized area given by the convex hull of the arrangement of points for both promising array structures; in (a) the value of directivity and in (b) the value of area occupied by the convex hull of the array,  selecting the optimal value of $d_\text{min}$. It is apparent in both graphics the remarkable directivity performance proximity for both structures and {respective} normalized areas with the increase in the number of antennas. The area occupied by the convex hull for small values of antennas are very similar{;} however, for the maximum directivity configuration and under a significant number of antenna elements, the UPA demonstrated a better performance covering a smaller region with the increasing of the antennas; besides, the points of the UHPA where the performance are significantly better is given when the hexagon formed by the array is perfect. Therefore, on can conjecture that both configurations have a very similar {directivity} performance. At the same time, for a significant number of antennas, the UPA surpasses the UHPA directivity {\it vs.} area performance tradeoff unless the hexagon formed by the UHPA is perfect.

Given the attained near directivity between UPA and UHPA, both geometries could be selected without losses in terms of directivity  {gain} and occupied area. Therefore, the structure selected for the optimization method in the next section is the UPA, due to the compactness in describing the antenna elements positioning over a rectangular coordinates system.

\section{Optimizing the Position of  UPA Elements} \label{sec:algorithm}
This section describes the proposed method to achieve high directivity using the regular planar array structure. This method provides reasonable solutions for the proposed optimization problem {of Eq.} \eqref{newproblem} with low computation effort considering the high-complexity operation of searching the geometric locus. 

\begin{figure}[htbp!]
\vspace{-5mm}
	\centering
\hspace{-4mm}	\includegraphics[width=.7\textwidth]{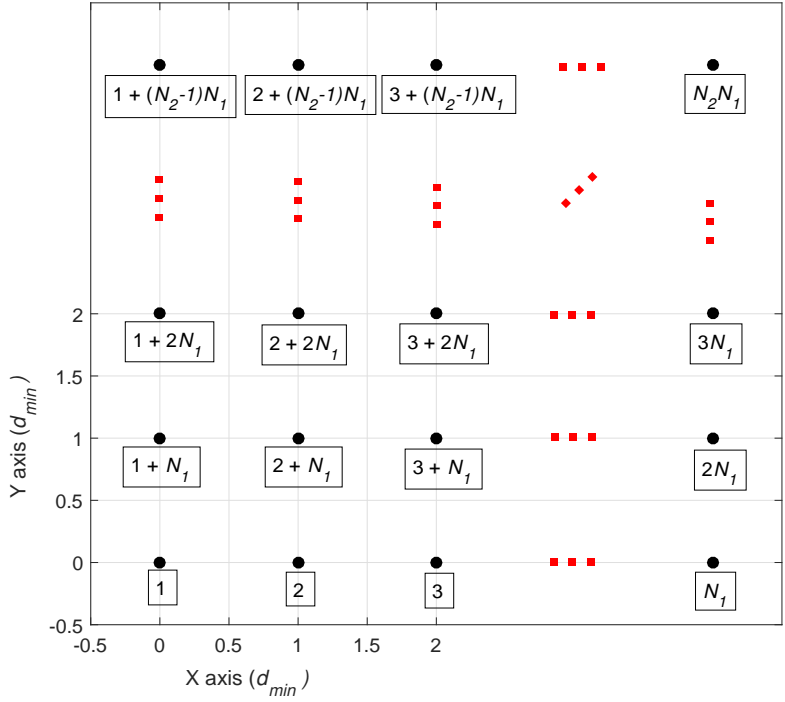}
	\vspace{-5mm}
	\caption{UPA elements disposition and labeling for $N_1 \times N_2$ antennas.} \label{fig:07}
\end{figure}

The proposed method consists in assuming a given number of waves ($k$), a desired direction ($\theta_0$,$\phi_0$), and a number of antennas ($N$), finding the optimal UPA on the desired plane defined by \eqref{zmn} aiming to attain a remarkable improvement in the directivity {in a} desired direction. The most critical parameter to be defined to achieve this goal is $d_\text{min}$. The following describes a procedure for finding this parameter for a generic UPA. Firstly,  we can consider the positioning of the antennas on the $ xy$ plane, considering a UPA with $N_1 \times N_2$ antenna elements regularly distributed, as in Fig. \ref{fig:07}.

From this antenna-elements distribution, one can define the distance between two generic elements, {\it i.e}, $d_{mn}$ for $m \neq n$, which is an important parameter on the use of \eqref{fdef}. The distance $d_{mn}$ can be formulated as:
\begin{align}
d_{mn} = \begin{cases}
\frac{| m - n | d_\text{min}}{N_1} &\text{if} \,\, \psi_1 = 0\\ 
| m - n |d_\text{min} \quad &\text{if} \,\, \psi_2 = 0  \quad \text{and} \quad | m - n | < N_1 \\ 
d_\text{min} \sqrt{ \psi_1^2 + \psi_2^2}  \quad \,\, &\text{c.c} 
\end{cases} \label{dmn2}
\end{align}
\text{where:}
\begin{align}
\begin{split}
&\psi_1 =  \text{mod}(m-1,N_1) - \text{mod}(n-1,N_1);  \quad \psi_2 =  \left \lfloor \frac{m - 1}{N_1} \right \rfloor - \left \lfloor \frac{n - 1}{N_1} \right \rfloor
\end{split}
\end{align}

To allocate the array on the desired plane without altering the distances, it is possible to use a rotation matrix considering the desired direction, defined by:
\begin{align}
&\boldsymbol{R}_{\boldsymbol{v}} =  \left[\begin{array}{ccc} 
\sin^2{\phi_0}\mu_{\gamma}+\cos{\gamma} & -\cos{\phi_0}\sin{\phi_0}\mu_{\gamma} & -\cos{\phi_0}\sin{\gamma}\\
-\sin{\phi_0}\cos{\phi_0}\mu_{\gamma} & \cos^2{\phi_0}\mu_{\gamma}+\cos{\gamma} &  - \sin{\phi_0}\sin{\gamma}\\
\cos{\phi_0}\sin{\gamma} & \sin{\phi_0}\sin{\gamma} & \cos{\gamma}
\end{array}\right ]
\label{transR}
\end{align}
where $\gamma = \acos{|\cos{\theta_0}|}$ and  $\mu_{\gamma} = (1-\cos{\gamma})$; notice that for $\theta_0 = 0$, the restriction \eqref{zmn} becomes the \textit{xy}-plane and this rotation matrix becomes an identity matrix.  The position matrix of the array on the desired plane can be expressed as:
\begin{align}\label{eq:Phat}
\widehat{\boldsymbol{P}} = \boldsymbol{P}\boldsymbol{R}_{\boldsymbol{v}}
\end{align}
{Besides, the} variable $z_{mn}$, which consist{s} in a combination of subtractions on the last column of $\widehat{\boldsymbol{P}}$ can be described as:
\begin{align}
z_{mn} = -(\psi_1 \cos{\phi_0}\sin{\gamma} + \psi_2 \sin{\phi_0}\sin{\gamma} )\label{zmn2}
\end{align}

Now, with the parameters $d_{mn}$ and $z_{mn}$ available, the OF analytical expression in \eqref{newproblem} can be {evaluated; however, this is a} {\it non-convex function} of one variable.  At this point, the directivity optimization problem can not be solved using convex optimization tools.  On the other hand, notice that finding the global optimum of this function, {either by exhaustive search or by quasi-optimal evolutionary heuristic methods}, the optimal UPA to be placed on the desired plane can be found, given the advantage of allocating the antenna-elements on the same plane, while finding the optimal general UPA in terms of directivity in that plane.

\vspace{3mm}
\noindent{{\bf Remark 1}: The general methodology is focused on finding the optimal position of each antenna element that maximizes the antenna array directivity. In this section, we have developed a solution confined into a specific plane, where the antenna-elements position on this plane resembles a sequence of equilateral triangles; hence, the idea of finding the optimal UPA antenna-elements placement in the such plane is described by the position matrix, Eq. \eqref{eq:Phat}, and solved applying the  simplified optimization problem, Eq. \eqref{newproblem}.}
\vspace{3mm}

\subsection{Successive Evaluation and Validation (SEV)}
\begin{figure}[!htbp]
\centering
\includegraphics[width=0.7\textwidth]{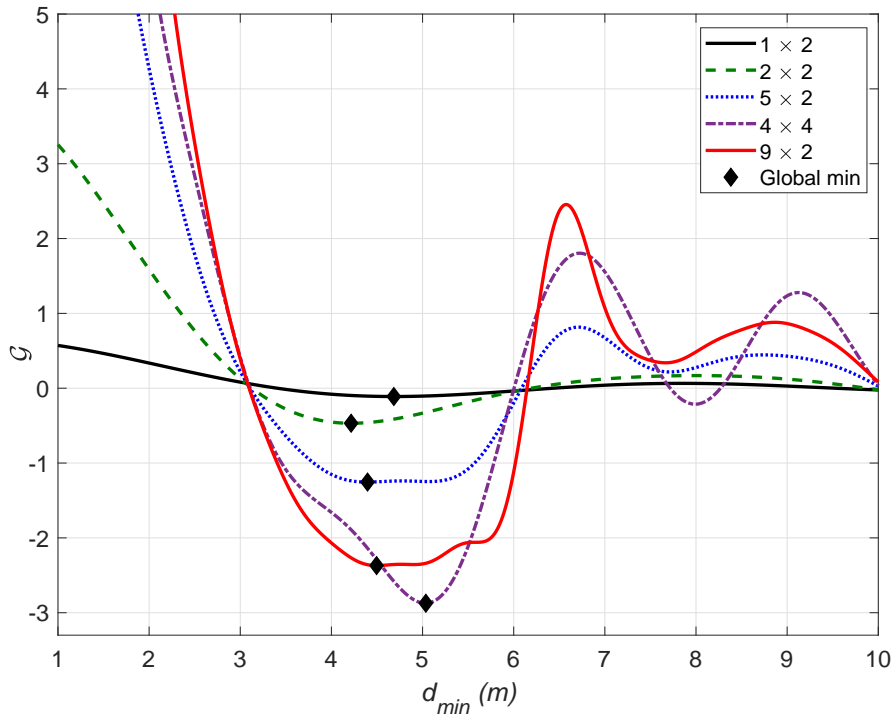}
\vspace{-3mm}	
\caption{{Values of $\mathcal{G}$ against  $d_\text{min}$ values for different configurations} of $N_1 \times N_2$}\label{fig:globalmin}
\end{figure}

With the $d_{mn}$ and $z_{mn}$ values, the {OF of the directivity optimization problem in} \eqref{newproblem} for the UPA can be expressed and therefore emerge{s} the necessity of find{ing} the optimal value {for} $d_\text{min}$ that enhances the array directivity.  {Based on} numerical results, we concluded that the first local minimum of the  {OF} in \eqref{newproblem} is also the global optimum. Fig. \ref{fig:globalmin} depicts the value of $\mathcal{G}$ for different values of $N_1$ and $N_2$. 
As suspected, for all analyzed  $N_1 \times N_2$ configurations, the first minimum local is also the {minimum} global.  {Taking} advantage of this aspect, we propose in Algorithm \ref{algo:SEV} the {\it successive evaluation and validation} (SEV) method, which basically {is a line search procedure, evaluating} the function {at the point ($d_0 = c$), with an increment closely to the origin} ($0^+$); then, an increment is computed again for $(d_1 =  d_0 + c$).  Hence, if the increment brought a {decreasing} of the OF, {\it i.e.}, $\mathcal{G}_{n+1} < \mathcal{G}_n$, then the process {is repeated;}  if not the point $d_n$ is declared the global optimum. 

\begin{algorithm}[!htbp]
\caption{SEV -- Successive Evaluation and Validation} \label{algo:SEV}
\begin{algorithmic}[1]
	\small
	\STATE \textbf{Input:}  $c$
	\STATE $n = 0$
	\STATE  $d_n = (n+1)c$
	\STATE $\mathcal{G}_{n} = - \sum\limits_{\substack{m,n=1 \\ m\neq n \\ n > m}}^N \mathcal{F}(d_{n}) \,\, ${: \,\, use \eqref{newproblem},	\eqref{dmn2},  \eqref{zmn2} with }$d_{n} = d_\text{min}$
	\WHILE{$\mathcal{G}_{n+1} - \mathcal{G}_{n} $ $\leq 0$}
	\STATE $n=n+1$
	\ENDWHILE
	\STATE ${d}_\text{min} = d_n$  	
	\STATE \textbf{Output:} ${d}_\text{min}^{\star}$
\end{algorithmic}
\end{algorithm}
A straightforward idea is behind the SEV algorithm: for a minimal input parameter $c$ value, the number of iterations increases; however, the precision of the solution improves; on the other hand, for large $c$ values, the number of iterations and precision can be reduced remarkably, and depending on how significant is such value, the first local minimum could be inadvertently skipped, losing the optimum solution.  Hence, the step parameter $c$ must be carefully selected to the SEV achieves a good precision-complexity tradeoff.

\subsection{Proposed Optimal Uniform Planar Array (OUPA)} \label{sec:OUPA}

The technique that allocates omnidirectional element-antennas arranged as uniform planar array (UPA) on a specific $z_{nm}$ plane, given by \eqref{zmn}, and using the SEV method to select the $d_\text{min}^{\star}$ is denominated hereafter {\it optimal uniform planar array} (OUPA). {We propose this straightforward, low-computational cost, quasi-optimal method} solution that significantly enhances the directivity in UPA antennas; besides, OUPA design requires few parameters: the carrier frequency/wave number ($f / \lambda$), the desired directivity angles ($\theta_0,\phi_0$) and the geometric UPA structure configuration, i.e., $N_1 \times N_2$.  A pseudocode for the OUPA technique is shown in Algorithm \ref{algo:OUPA}.

\begin{algorithm}[!htbp]
\caption{OUPA -- Optimal Uniform Planar Array} \label{algo:OUPA}
\begin{algorithmic}[1]
	\small
	\STATE \textbf{Input:}  $f/ \lambda$, $\theta_0$, $\phi_0$, $N_1$, $N_2$\vspace{0.1cm}
	\STATE Calculate all possible $d_{mn}$ values using \eqref{dmn2}
	\STATE Calculate all possible $z_{mn}$ values using \eqref{zmn2}
	\STATE Evaluate $\mathcal{G}$ in \eqref{newproblem}  using the values of $d_{mn}$ and $z_{mn}$ 
    \STATE Determine $d_\text{min}^\star$ via SEV method (Algorithm \ref{algo:SEV})
	\STATE Determine the UPA matrix of position $\boldsymbol{P}$ using  $d_\text{min}^\star$ via \eqref{eq:P} 
	\STATE Final array position evaluated by $\widehat{\boldsymbol{P}}(\theta_0 , \phi_0) = \boldsymbol{P}\boldsymbol{R}_v$,   eq. \eqref{eq:Phat}  \vspace{0.1cm}
		
	\STATE \textbf{Output:} $\widehat{\boldsymbol{P}}$ 	
\end{algorithmic}
\end{algorithm}

\vspace{3mm}
\noindent{{\bf Remark 2}: The {OUPA method} exploits the OF features,  Eq. \eqref{fdef}, in the simplified optimization formulation, Eq. \eqref{newproblem}, while for generic $N_1\times N_2$ antennas configuration, a practical quasi-optimal solution is developed in the next section (specifically, subsection \ref{sec: GA}) based on the evolutionary heuristic optimization approach. The array directivity subject to antenna-elements positioning has been formulated and solved for various elements in the range $N \in \{4;\, 36\}$.}

\section{Numerical Results} \label{sec:numerical}
Different OUPA performance aspects are numerically evaluated, including a) directivity performance, b) occupied area, c) comparison with other directivity optimization methods. The general parameters for the omnidirectional directivity optimization scenario deployed throughout this section are summarized in Table \ref{tab:scen}.

\begin{table}[!htbp]
\centering
\caption{General simulation setup}
\begin{tabular}{ll}
\hline
\bf Parameter & \bf Adopted Values \\
\hline
Angle of Departure & ($\theta_0$ ,$\phi_0$) = ($\frac{\pi}{4}$, $\frac{\pi}{4}$) \\	
Element Amplitude & $A_n = 1$ for all $n$.\\
Element Phase & $\alpha_n$ = 0 for all $n$. \\
Omni-directional Scenario & $u=0$ and $v=1$\\
\hline
\end{tabular}\label{tab:scen}
\end{table}

\subsection{{OUPA Directivity Performance}} \label{sec:OUPA_dir}
Numerical simulations have been conducted considering different values of $N_1\times N_2$ antenna arrangements.  For each configuration, the value of $d_\text{min}^\star$ is obtained using the SEV procedure, and then the relation directivity-area occupied by the array is examined. The goal is understanding how the occupied area impacts the planar array directivity. The occupied array area is an essential parameter since compact antenna arrays are far more helpful.  Table \ref{tab:II} depicts the selected parameter values deployed in the simulations.

\begin{table}[!htbp]
\centering
\caption{{Simulation setup for OUPA directivity-area evaluation}}
\begin{tabular}{ll}
\hline
\bf Parameter & \bf Adopted Values  \\ 
\hline
Frequency / Wave Number & $5$ GHz / $k \approx 104.8$ m$^{-1}$ \\
{SEV parameter (increment)} & $c = 10^{-3}$ \\
\hline
\vspace{-0.1cm} & \vspace{-0.1cm} \\ 
\multicolumn{2}{c}{\bf Number of $N = N_1 \times N_2$ antennas-elements}\\			
\hline
{\# Vertical elements } & $N_1 = [2, 3, 4, \dots 50] $\\ 
{\# Horizontal elements} & $N_2 = [2, 3, 4, \dots 10] $\\ 
\hline
\end{tabular}\label{tab:II}
\end{table}

\begin{figure*}[htbp!]
\centering
\includegraphics[trim={12mm 5mm 2mm 12mm},clip,width=0.51\textwidth]{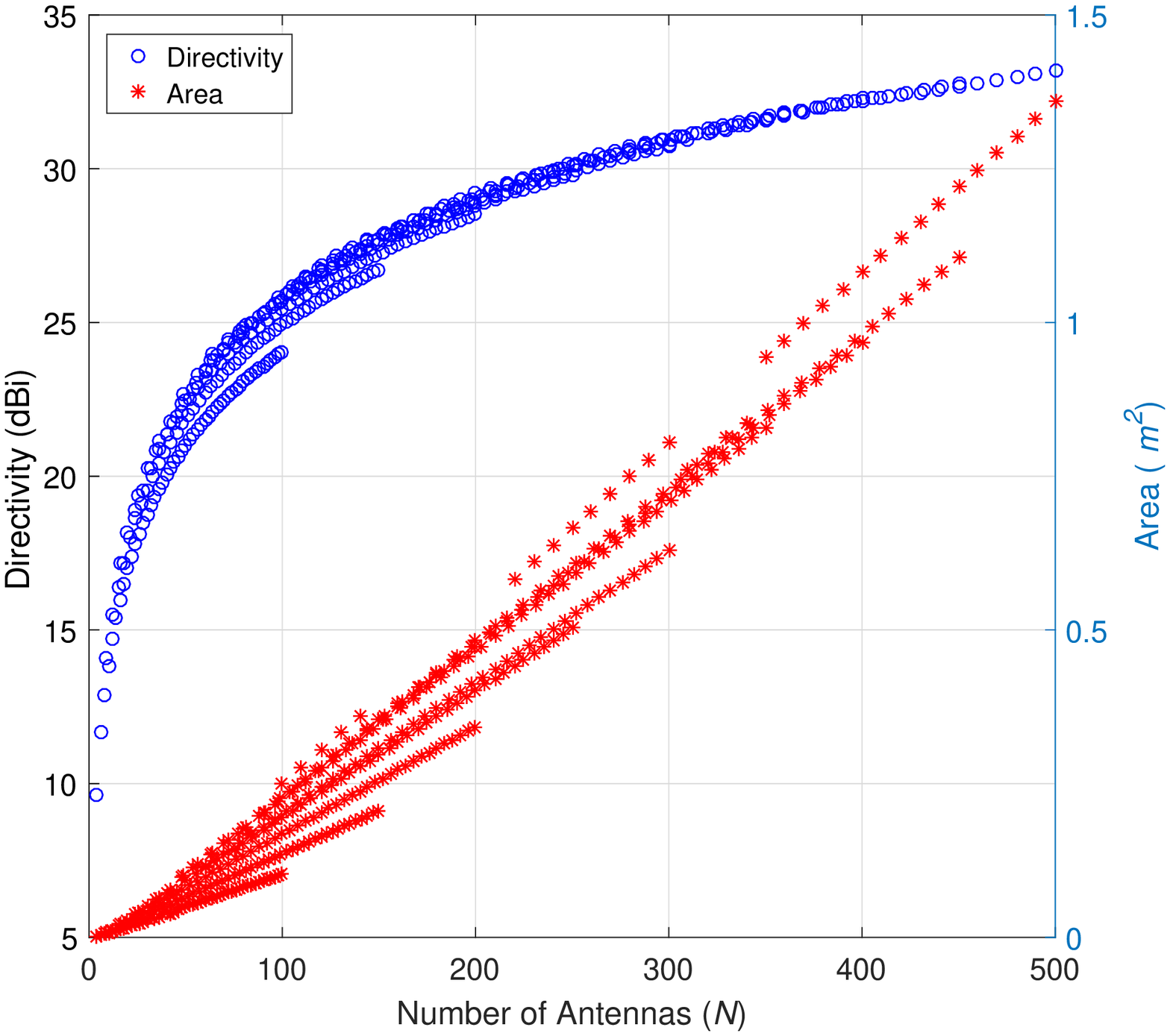}
\includegraphics[trim={10mm 5mm 15mm 13mm},clip,width=0.48\textwidth]{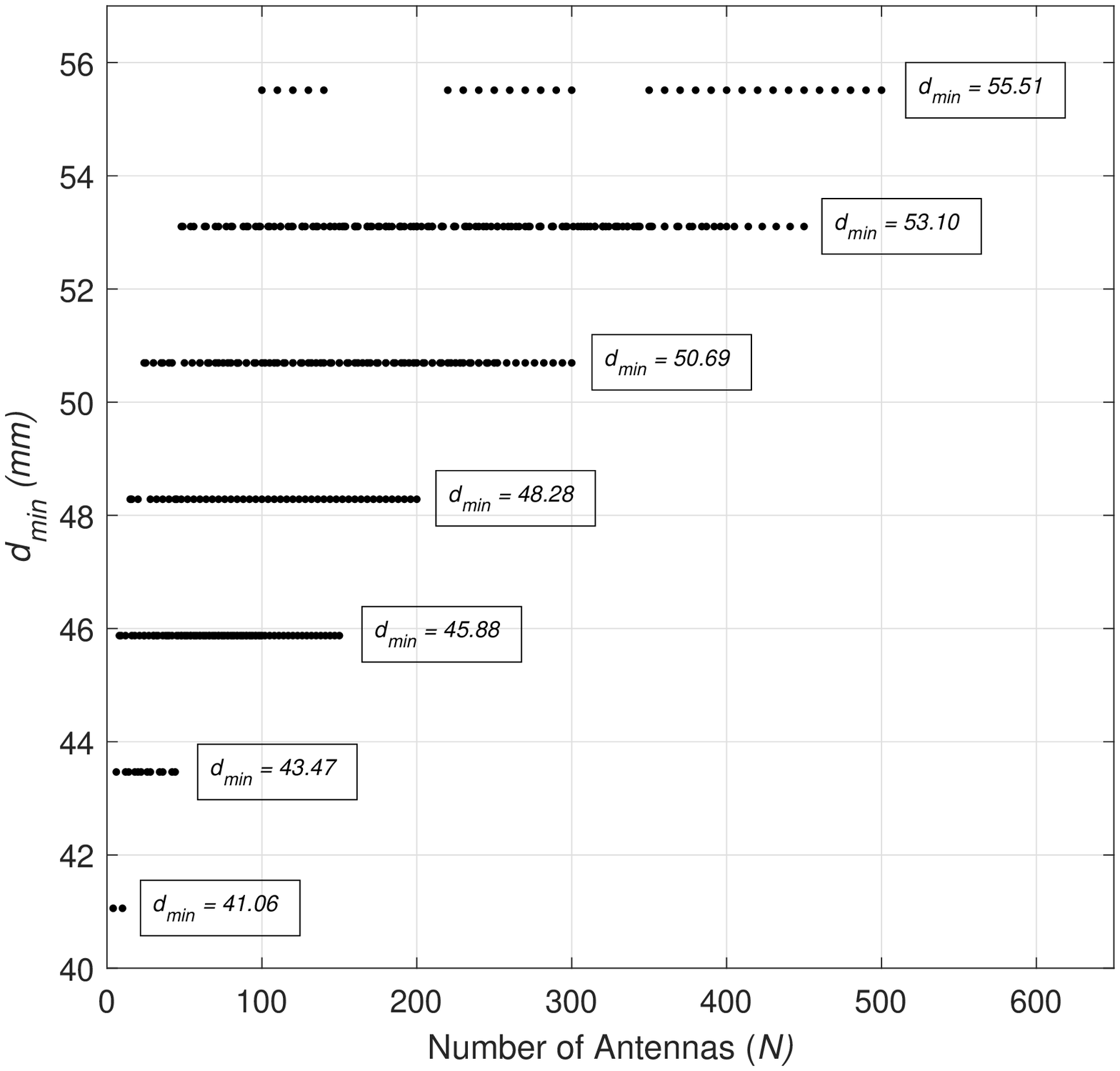}
\caption{{{\bf a}) Directivity and array Area values for different $N=N_1 \times N_2$ element-antennas configurations; {\bf b}) correspondent $d_{\min}^\star$ values obtained via SEV procedure for different configurations of $N=N_1\times N_2$.}}
\label{fig:14}
\end{figure*}
 In Fig. \ref{fig:14}, the values of directivity, {array area and correspondent $d_\text{min}^\star$} are depicted for a wide range of $N$ element-antennas values. As expected, increasing $N$, the directivity and array area are increased proportionally. {Since the array area increment is not desired, the directivity-area tradeoff must be carefully evaluated.  The correspondent $d_\text{min}^\star$ values found by simulations using SEV procedure} are illustrated in Fig. \ref{fig:14}.b.  The most important feature of this plot is that the optimal value of $d_\text{min}$ has a discrete distribution {regarding} $N_1$ and $N_2$; besides, the difference between the levels of optimum values is constant, which is very interesting and can be used to reduce significantly the search space of the optimal value. Another essential characteristic is that the increase in $N$ not necessarily increases $d_\text{min}^\star$ value. Finally, the $N_1$ and $N_2$ values impact the optimal $d_\text{min}$ value; hence, this effect is analyzed in more details in the next section.

\begin{figure}[htbp!]
\vspace{-4mm}
\centering
\includegraphics[trim={12mm 3mm 2mm 12mm},clip,width=0.6\textwidth]{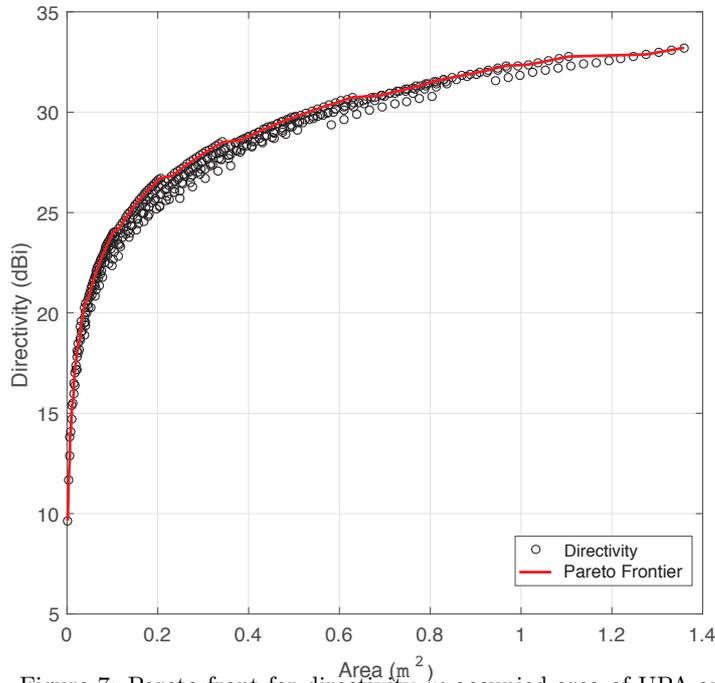}
\vspace{-6mm}
\caption{{Pareto front for directivity {\it vs} occupied area of UPA arrays.}}
\label{fig:pareto}
\end{figure}

To gain a deep understanding of the Area and Directivity dependence, Fig. \ref{fig:pareto} depicts directivity values against array area for different configurations of $N=N_1\times N_2$. Observing the Pareto frontier among conflicting directivity {\it vs} occupied area by the UPA array is possible. As expected, it is necessary to increase the array area to achieve higher directivity values. The Pareto front establishes the optimal trade-off between both conflicting parameters; possible solutions (points) above this frontier are unfeasible for such an array structure, and the points below are sub-optimal in the directivity-area tradeoff.

\subsection{Directivity {\it vs} Area under Different $N_1 \times N_2$ Arrangements}
Due to constructive aspects and integer combination of $N_1$ and $N_2$, the structure realization for a given area is constrained, given that the total number of antennas is a multiplication of two integers, i.e., $N_1 \times N_2 = N$. Moreover, most of the analyses available in the literature focus on the UPA structures, with squared $N_1 = N_2$ antenna arrangements. In this work, we also have analyzed rectangular $N_1 \neq N_2$ antenna arrangements.

\begin{figure}[!htbp]
\centering
\hspace{-6mm}	
\includegraphics[width=.656\linewidth]{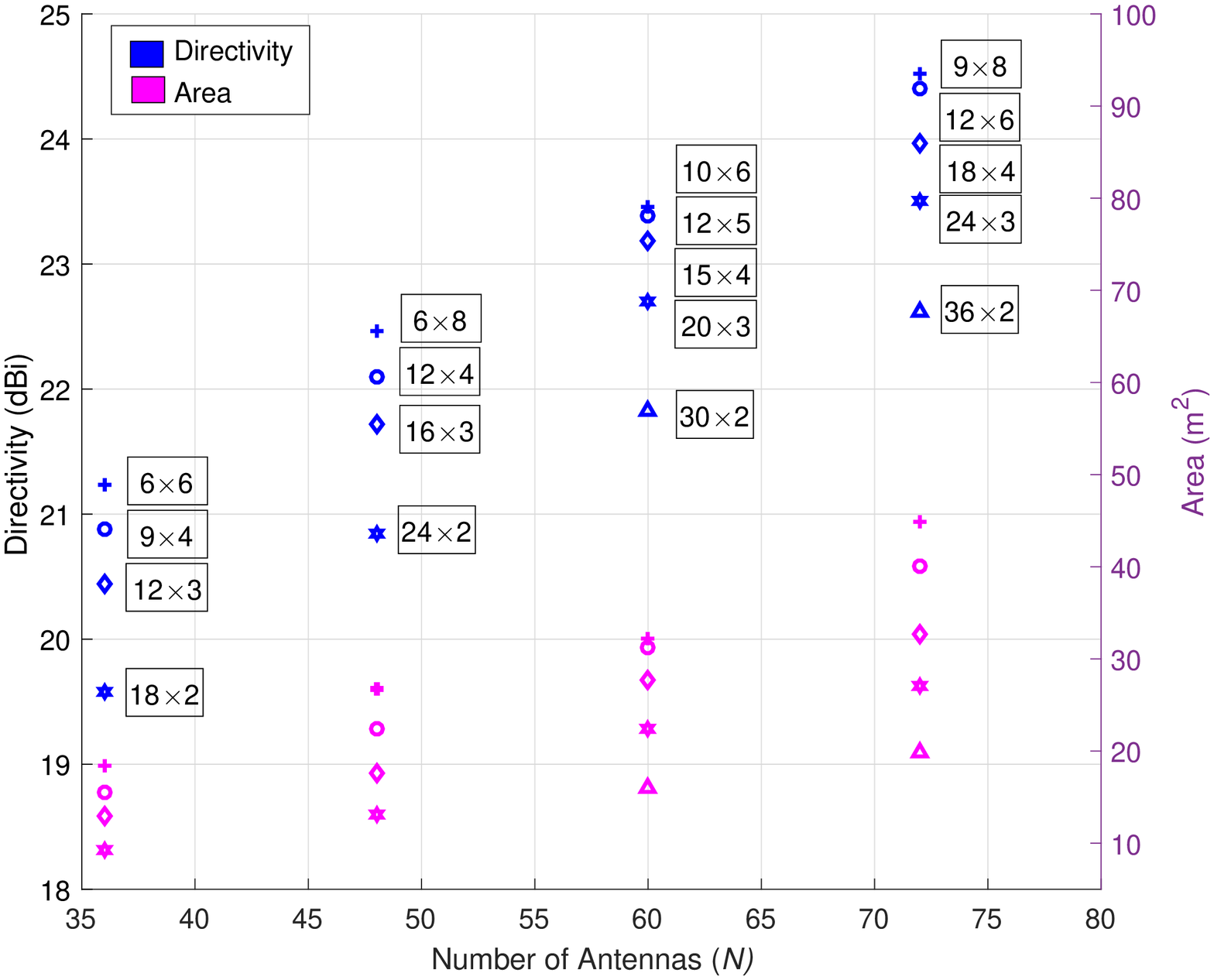}\,\,\,\,
\includegraphics[width=.34\linewidth]{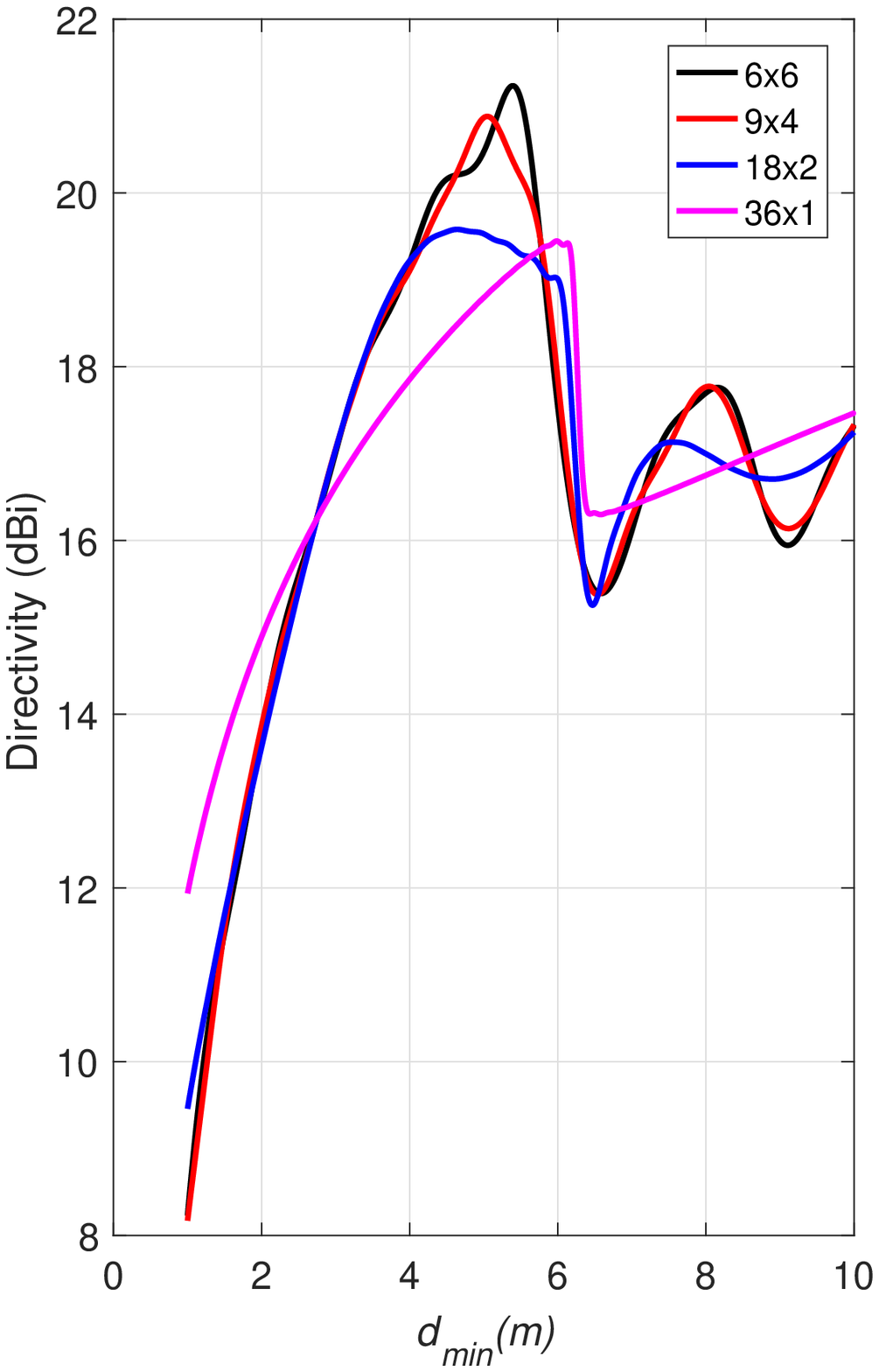}\\
\small \hspace{20mm} (a)    \hspace{60mm} \small (b)
\vspace{-2mm}
\caption{{\bf a}) Directivity and Area {\it vs.} $N$ for different configurations with the same number of antennas $N=[36; 48; 60; 72]$. \quad {\bf b}) Directivity {\it vs} $d_{\min}$ fo four possible $N_1 \times N_2$ arrangements in an UPA with $N=36$ antennas.}
\label{fig:15}
\end{figure}

To investigate the impact on the directivity, four different UPA structures with $N=[36; 48; 60; 72]$   {antennas were compared}.  Fig. \ref{fig:15}.a) depicts possible {achievable directivity values} and respective UPA normalized area ({assuming} $k=1$) for {possible $N_1$ and $N_2$ combinations}. The directivity value and area occupied by $N$ antennas are compared considering different UPA configurations, each column {representing} the same number of antennas; the markers on the same column indicate specific configurations of $N_1 \times N_2$. As indicated in Fig. \ref{fig:15}.b), decreasing the difference between $N_1$ and $N_2$ results in better values of maximum attainable directivity\footnote{In Fig. \ref{fig:15}.b), the variation of $d_\text{min}$ causes an impact on the value of directivity. It is possible to notice that better directivity values are achieved when $N_1 = N_2$. Indeed, these numerical results indicate that the smaller difference between $N_1$ and $N_2$ better is the array directivity.} as a function of $d_{\min}$. However, the area occupied by the UPA also increases. {On the other hand, choosing} configurations with a great disparity among $N_1$ and $N_2$ implies a substantial directivity reduction, being possible to attain a similar performance, or even be surpassed, when compared with the same UPA structure but with a smaller number of antennas,  for instance, $36\times 2$ against $6\times8$, in which, despite the great difference of $24$ antennas, the directivity are quite similar in both array arrangements.
  
As the focus of this work is the directivity enhancement and aiming at exploiting the UPA structure, the difference between $N_1$ and $N_2$ {must be reduced; therefore, the choice for $N_1$ and $N_2$ should respect the following condition:}
\begin{align}
|N_1 - N_2 | \,\, = \,\, \left\{ {\begin{array}{*{20}{ll}}
0 &  \text{squared (UPA)}\\
1 & \text{rectangular (UPA)}\\
\end{array}} 
\right.
\label{n1n2}	
\end{align}

\subsection{{Genetic Algorithm solving Non-Convex Directivity Problem}} \label{sec: GA}
 Since the original directivity optimization is a non-convex problem,  in this subsection, the performance of the genetic algorithm (GA) in solving the omnidirectional uniform array directivity problem is evaluated. The pseudocode for the GA deployed to find the enhanced directivity solutions is presented in Algorithm \ref{algo:GA}.  Besides, the  GA parameter values deployed in this context are summarized in Table \ref{tab:GA}.

\begin{algorithm}[!htbp]
\caption{{GA-UPA -- GA} Optimizer for Omnidirectional Uniform Antenna Array} \label{algo:GA}
\begin{algorithmic}[1]
\small
\STATE \textbf{Input:}  $n_{\text{vars}}$, $c_{f}$, $g_\text{max}$, $M_\text{Gaussian}$, $p_{\text{size}}$ 
\STATE  \textit{Population} $\leftarrow$ InitializePopulation($p_{\text{size}}$,$n_{\text{vars}}$)
		\STATE $\boldsymbol{S} = \mathcal{G}(\textit{Population})$
		\STATE $\boldsymbol{s}_{\text{best}}$ $\leftarrow \boldsymbol{s}$ \textit{which results in the minimum value of}  $\mathcal{G}$($\boldsymbol{s}$)
		\WHILE {$\# \text{\textit{ number of Generation}} < g_\text{max}$}
		\STATE \textit{Parents} $\leftarrow$ SelectParents(\textit{Population},$p_{\text{size}}$)
		\STATE \textit{Children} $\leftarrow$ $\emptyset$
		\FOR{[\textit{Parent}$_1$,\textit{Parent}$_2$] $\in$ \textit{Parents}}
		\STATE [\textit{Children}$_1$,\textit{Children}$_2$] $\leftarrow$ Crossover(\textit{Parent}$_1$,\textit{Parent}$_2$,$c_{f}$)
		\STATE \textit{Children} $\leftarrow$ $M_{\text{Gaussian}}$(\textit{Child}$_1$)
		\STATE \textit{Children} $\leftarrow$ $M_{\text{Gaussian}}$(\textit{Child}$_2$)
		\ENDFOR
		\STATE $\boldsymbol{S} = \mathcal{G}(\textit{Children})$
		\STATE $\boldsymbol{s}_{\text{best}}$ $\leftarrow \boldsymbol{s}$ \textit{which results in the minimum value of}  $\mathcal{G}$($\boldsymbol{s}$)
		\STATE \textit{Population} $\leftarrow$ Replace(\textit{Population}, \textit{Children})
		\ENDWHILE 		
		\STATE \textbf{Output:} ${\boldsymbol{s}}_{\text{best}}$
	\end{algorithmic}
\end{algorithm}

\begin{table}[!htbp]
\centering
	\caption{Simulation Setup used in the {GA-UPA directivity}}
	\begin{tabular}{ll}
		\hline
		\bf Parameter & \bf Adopted Values \\
		\hline
		Wave Number / Wavelength & $k=1$ $[m^{-1}]$ / $\lambda = 2\pi$ $[m]$\\
		\# antenna-elements  & $N\in \{4,5,6,7,8,9\}$ \\
		Elements placement bounds & $[x_{\max},\, y_{\max},\, z_{\max}] = [5, \, 5, \, z_{\max}]$  \\
		Objective function (OF) & eq. \eqref{newproblem} \\
		Number of Variables, ($n_{\text{vars}}$) & $2N$ \\
		Crossover Fraction ($c_{f}$) & 0.7\\
		Max. \# Generations ($g_{\text{max}}$) & $40$\\
		Mutation ($M_{\text{Gaussian}}$) & Gaussian \\
		Initial Population & Feasible and Random  \\
		{Population Size} ($p_{\text{size}}$) & {$200 N$} \\
		{Population Type} & {Double Vector} \\
		Search space &  $[0,\, x_{\max}]$; \, \, $[0, \, y_{\max}]$\\ 
		\hline
	\end{tabular}\label{tab:GA}
\end{table}

\vspace{3mm}
\noindent{{\bf Remark 3}: GA has been selected as a powerful, well-established evolutionary technique to solve non-convex optimization problems, among other also well-known techniques, such as particle swarm optimization (PSO), ant colony optimization (ACO), artificial bee colony algorithm (BCA), and other evolutionary algorithms (EA) optimization. GA or EA applies the natural evolution principles to find an optimal local solution. In GA, the problem is encoded in a series of bit strings manipulated by the algorithm; the decision variables and problem functions are deployed directly. The main {\it drawbacks} of an EA are: a) it is much slower\footnote{often by factors of a hundred times.} than alternatives such as the gradient-based and Simplex methods; b) as problem size scales up, {\it e.g.}, ten to a hundred or a thousand decision variables, an EA is often overwhelmed by the dimensionality of the problem, being unable to find a solution close to a locally optimal solution; c) a solution is acceptable only in comparison to other, previously discovered solutions; indeed, an EA has no concept of an {\it optimal solution}, or any way to test whether a given solution is optimal, even locally optimal; d) An EA never really knows when to stop, since it does not know whether a given solution is optimal. Finally, EAs usually finish running manually by the user, or by a predetermined limit on the number of iterations.}
\vspace{3mm}

\begin{figure}[!htbp]
\centering
\subfigure[$N=6$ antennas]{\includegraphics[width=.53\textwidth]{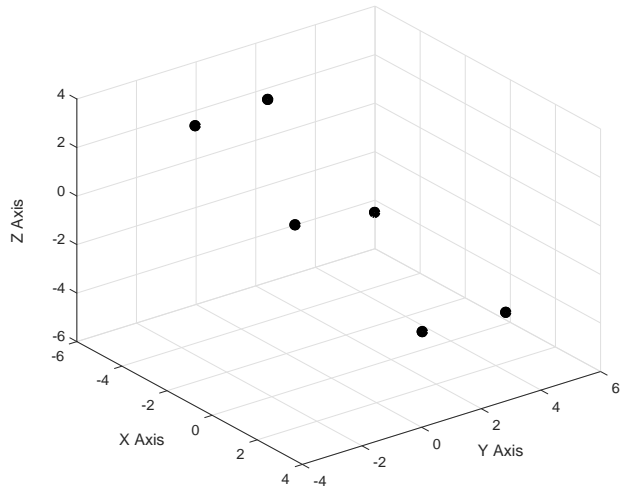}} 
\subfigure[$N=6$, $xy$-plane]{\includegraphics[width=.44\textwidth]{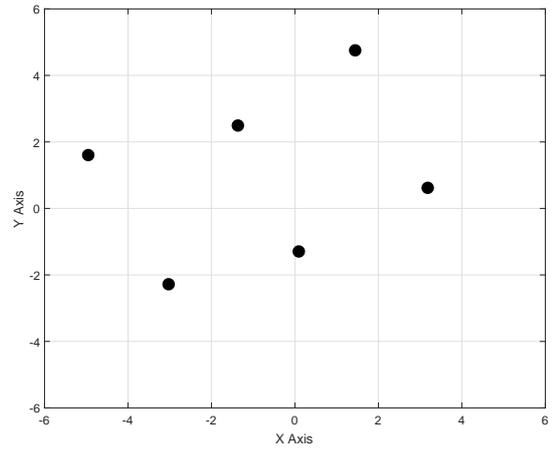}}\\
\subfigure[$N=8$]{\includegraphics[width=.53\textwidth]{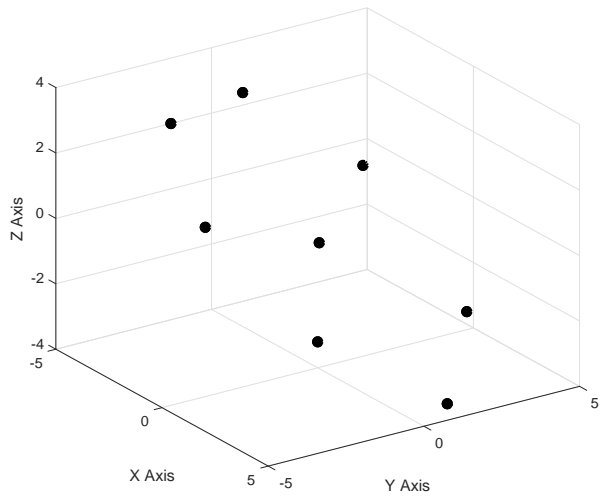}} 
\subfigure[$N=8$, $xy$-plane]{\includegraphics[width=.44\textwidth]{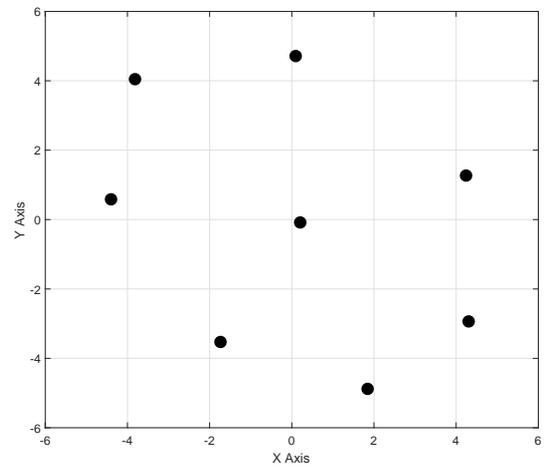}}
\caption{{GA solution for $N = 6,\, 8$ UPA antennas: ({\bf left}) final element-antennas position; ({\bf right}) projection onto the $xy$-plane.}}
\label{fig:10}
\end{figure}

Fig. \ref{fig:10} depicts the solution of problem {eq.} \eqref{newproblem} found by the GA for $N=6, 7$ and $8$ antennas. When the solution is visualized in the 3-D plots, it is possible to observe the plane that restrains the solution and how the solutions found strive to attain a significant number of {\it equal} Euclidean {\it distances}. However, when analyzing the projection of the solution in the $xy$-plane, it is possible to verify the distortion introduced by the term $z_{nm}$, eq. \eqref{zmn}. 

\begin{figure}[!htbp]
\vspace{-3mm}
\normalsize
\centering
\subfigure[$1^\text{th}$ Generation]{\includegraphics[width=.35\textwidth]{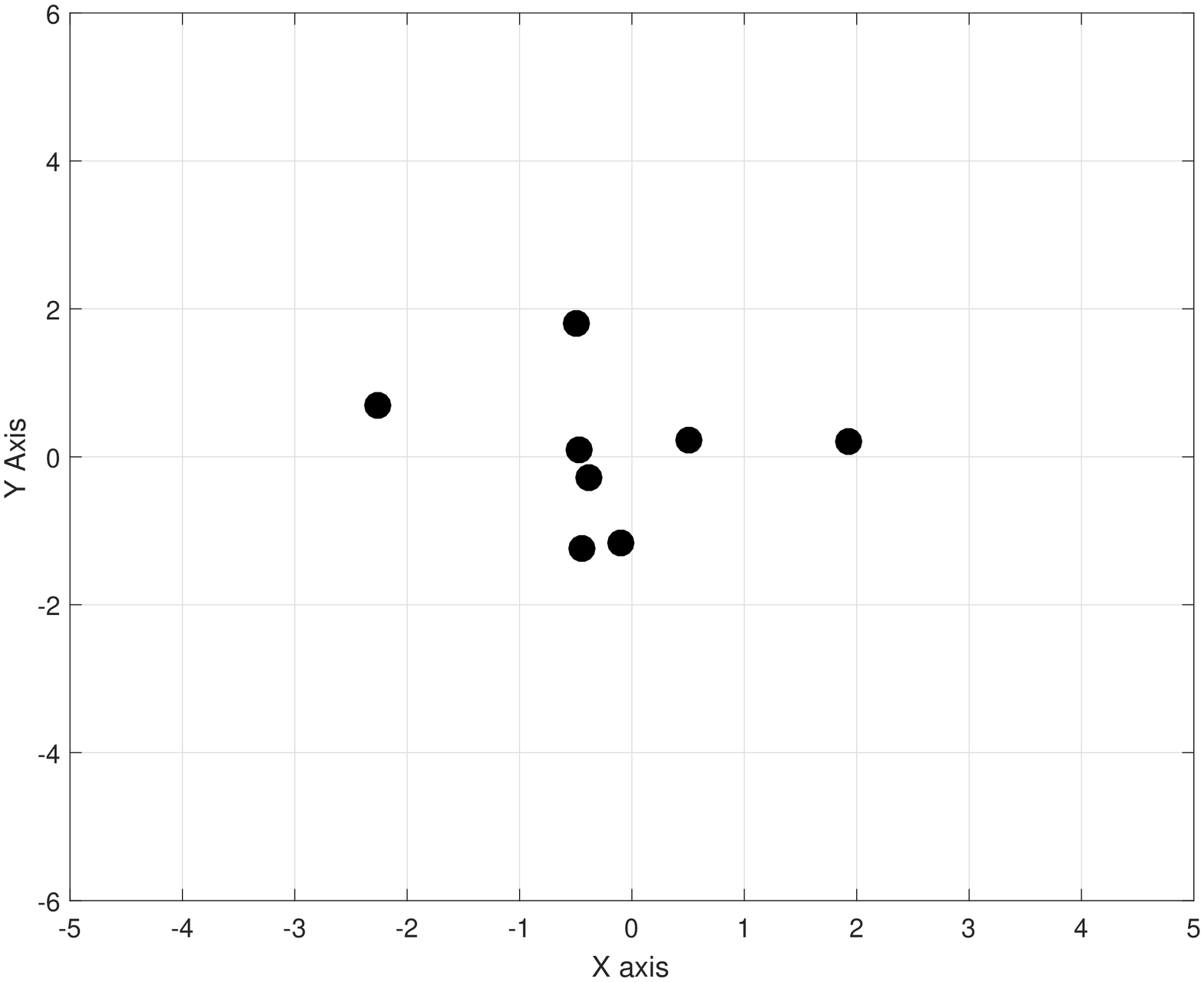}}
\subfigure[$1^\text{th}$ Generation]{\includegraphics[width=.62\textwidth]{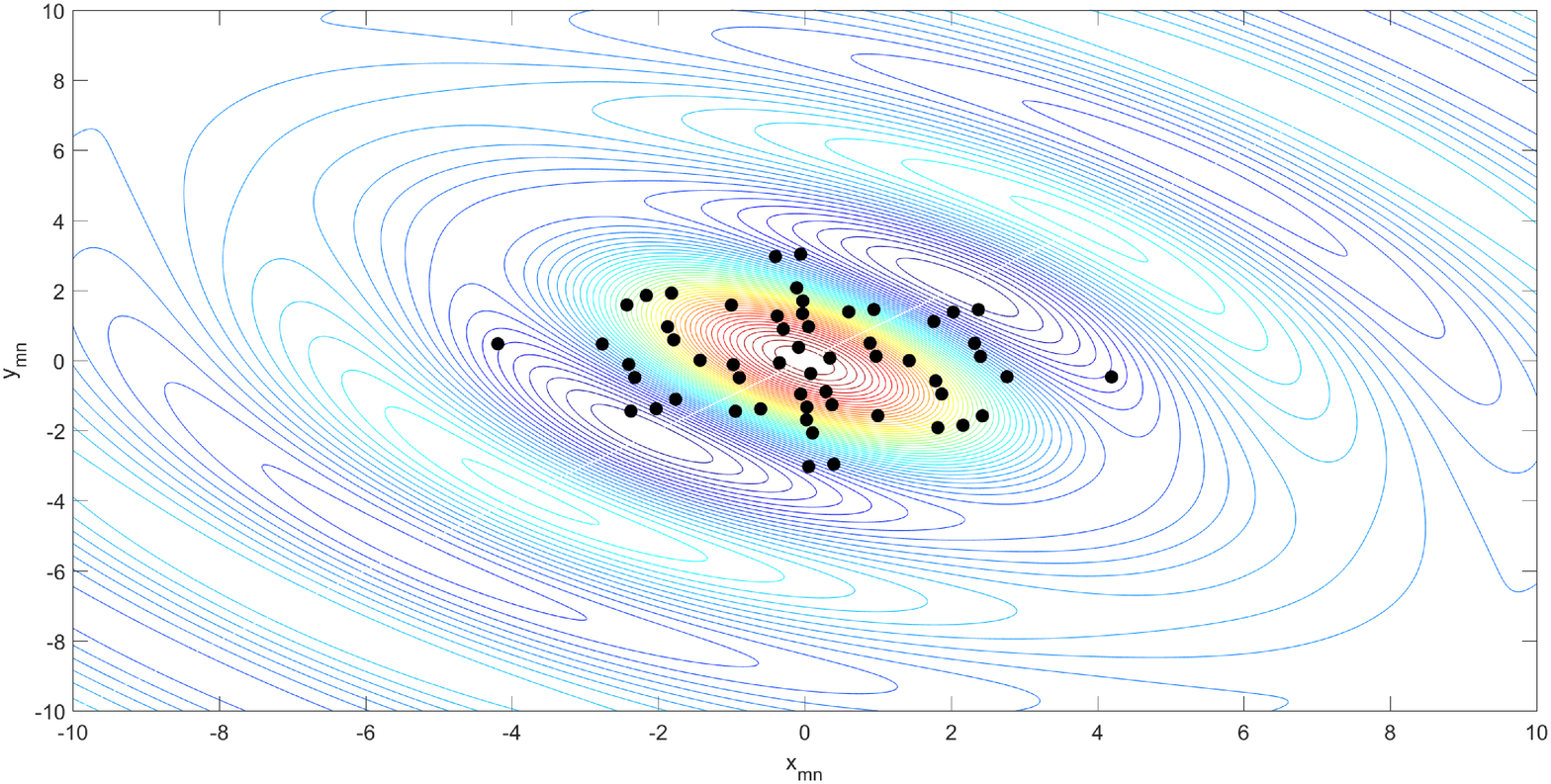}} 
\subfigure[{$35^\text{th}$} Generation]{\includegraphics[width=.35\textwidth]{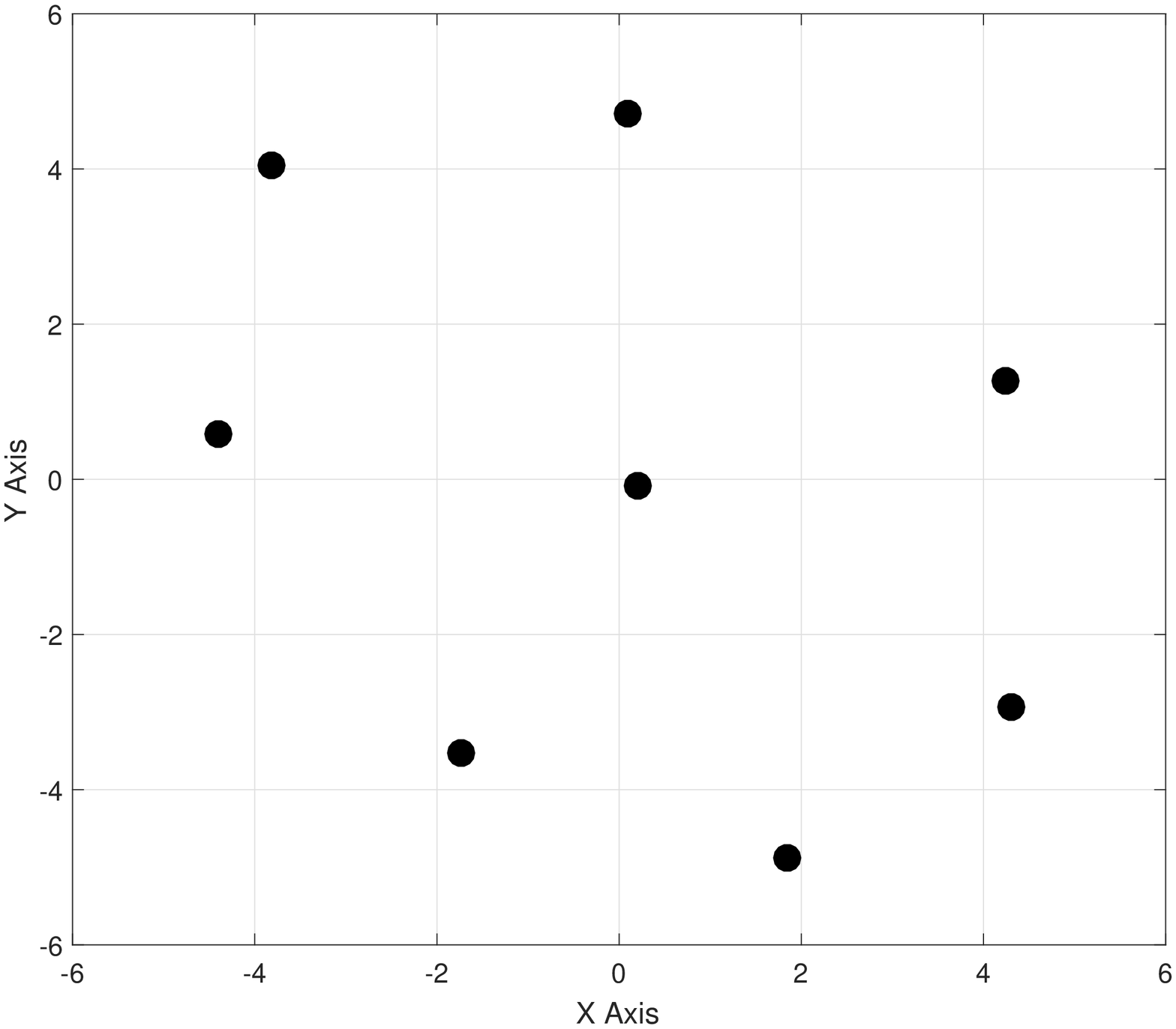}}
\subfigure[{$35^\text{th}$} Generation]{\includegraphics[width=.64\textwidth]{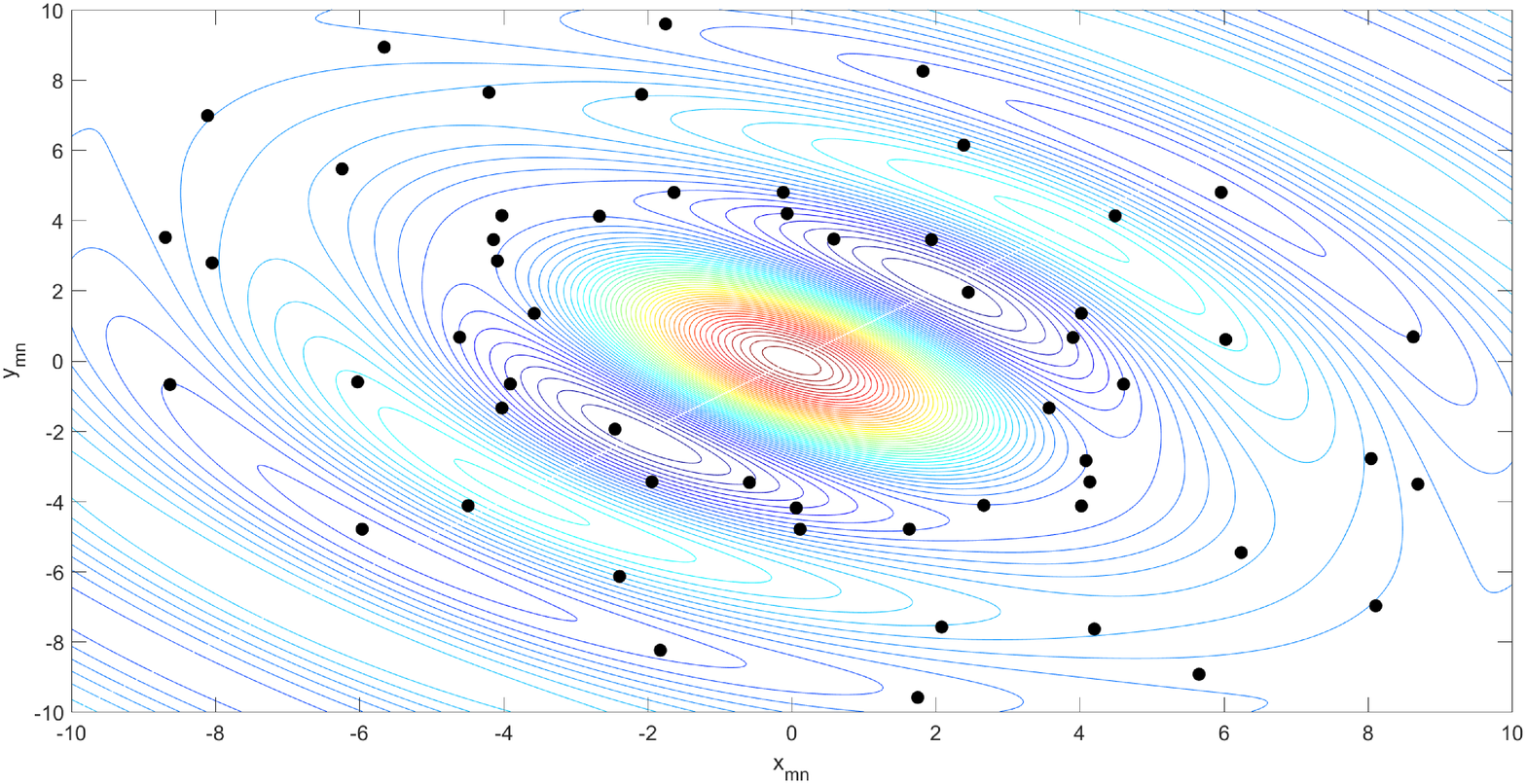}}
\caption{{Evolution of GA-UPA solutions for $N=8$ antennas:  ({\bf left}) antenna-element position;  ({\bf right}) distribution of ($x_{mn}$,$y_{mn})$ and contour lines for the objective function $\mathcal{F}$ of eq. \eqref{fdef}.}} 
\label{fig:11}
\end{figure}

The evolution of antenna-element position through the GA-UPA generations is depicted in Fig. \ref{fig:11}; the left plots show the placement of the coordinates ($\boldsymbol{x},\boldsymbol{y}$) found by the GA solution and {the right plots indicate} the distribution of the {difference-coordinate points} ($x_{mn}$,$y_{mn}$), {\it i.e.}, all possible differences between the correspondent coordinates ($\boldsymbol{x},\boldsymbol{y}$){, together with the contour plots of the objective function $\mathcal{F}$ defined in \eqref{fdef}. Notice that in the right plots,} the number of points on each plot is given by the combinations of all possible {differences $(x_{mn}$,$y_{mn})$,  which result in $ N(N-1)$ points}. Another interpretation for these coordinates points is that they represent the terms of each summation in $\mathcal{G}$, {eq. \eqref{newproblem}}. 
The evolution of the antenna-element positions through the GA generations is given in the \textit{xy}-plane projection. Moreover, it is possible to observe the evolution of the coordinates points ($x_{mn}$, $y_{mn}$); this evolution leads to an arrangement where all {allocated points are concentrated in sites of small values of $\mathcal{F}$, regarding} the range of all the possible values.

The selection of the {difference-coordinate points} ($x_{mn}, y_{mn}$) can not be performed freely. {Indeed, these points} result in  Euclidean distances constraints, which in some cases {did not physically} feasible; therefore, the obvious solution consisting in allocating all points {to} the same minimum of $\mathcal{F}$ is not possible due to the implication that all Euclidean distances must be the same. Indeed, the modification of position in one element of the antenna implies in the allocation of $N-1$ points; hence, the allocation of these points must be resolved simultaneously.

\vspace{3mm}
\noindent{{\bf Remark 4}: The GA solution (Fig. \ref{fig:11}) converges to a geometric figure similar to the points on a {\it regular triangular tiling} (RTT)}; the only reason for the convergence is not exactly an RTT geometry is found examining plane constrain $z_{mn}$, eq. \eqref{zmn}; hence, for $\theta_0 = \phi_0 = 0$, the solution will be given by a set of points allocated on an RTT.  This arrangement presents symmetry which imposes that all possible distances {can be reduced remarkably} {by repetition of points regularity. As a result, the difference-coordinate} points ($x_{mn}$,$y_{mn}$) can be distributed exploiting {such standard, selecting the feasible points corresponding} to small values of $\mathcal{F}$ and using the repetition to allocate more points.

\vspace{4mm}
\noindent\textbf{\textit{Comparison with the UCA and  ULA}}:
{The values of directivity found with GA-UPA} is compared with the two classical steering vector beamforming: a) the {\it uniform circular array} (UCA), and b) the {\it uniform linear array} (ULA). The directivity using steering vector consists of changing each antenna element's phase, aiming to minimize the array factor in \eqref{arraygeral}, depending on the position of each element in a well-defined geometric structure. For the ULA and UCA the {\it steering vector} is given, respectively, by:
\begin{align}
&\alpha_{n}^{\textsc{ula}} = -d n\cos{\theta_0} \, \, \hat{k}\\
&\alpha_n^{\textsc{uca}} = -r\sin{\gamma_n}\sin{\theta_0}\cos{\phi_0} \, \, \hat{i} - r\cos{\gamma_n}\sin{\theta_0}\sin{\phi_0}\, \, \hat{j}\notag
\end{align}
where: $\gamma_n = \frac{(n-1)2\pi}{N}$, and $d$ is the regular distance between adjacent element. In this formulation,  ${r}$ is the ratio of the circle that contains all the points of the UCA. This ratio was considered so that the minimal distance between the antenna's element was defined as half carrier wavelength,  $\frac{\lambda}{2}$.

\begin{figure}[!htbp]
	\centering
	\includegraphics[trim={1mm 2mm 7mm 12mm},clip,width=.35\linewidth]{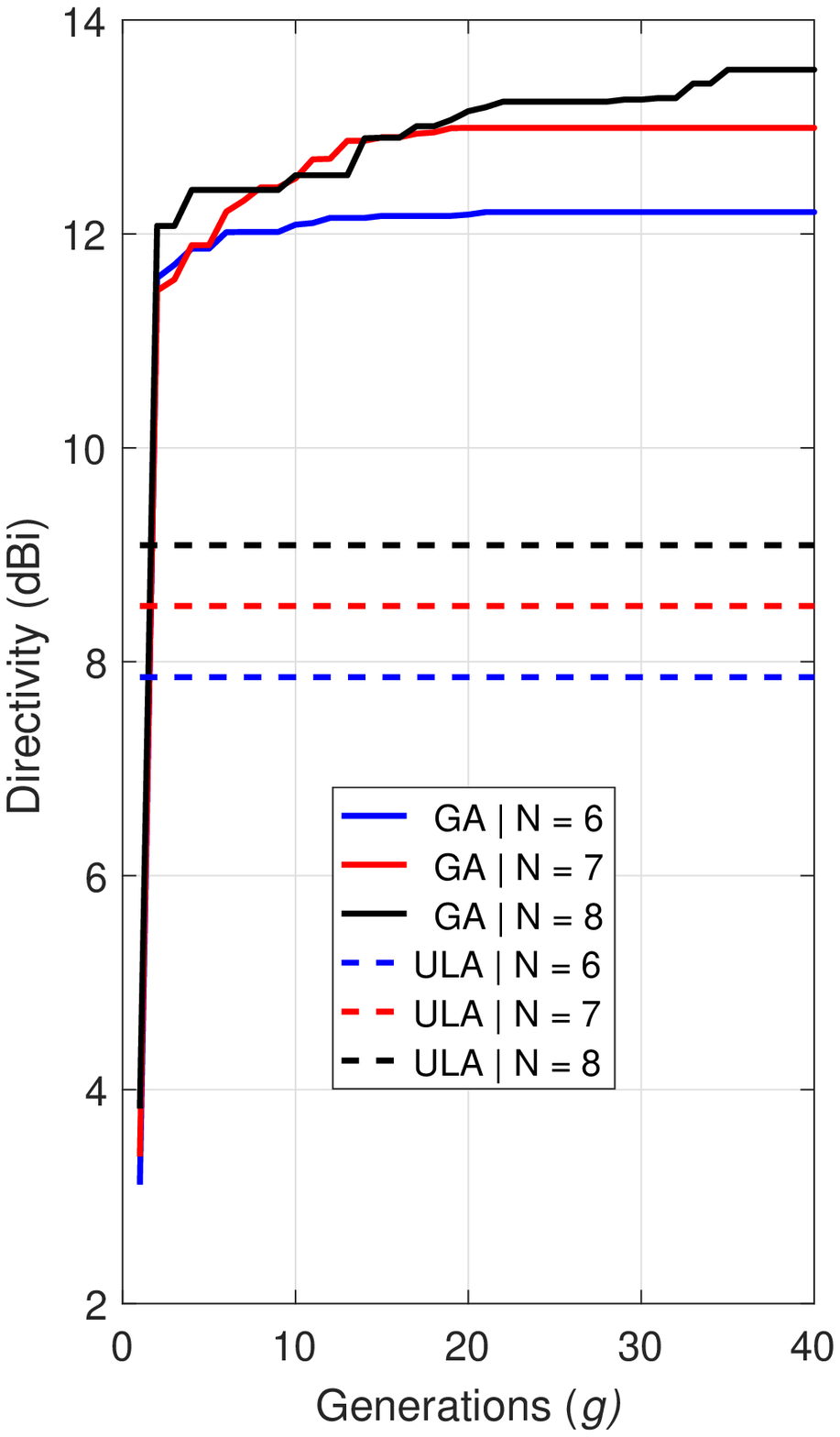}
	\includegraphics[trim={3mm 8mm 10mm 12mm},clip,width=.415\linewidth]{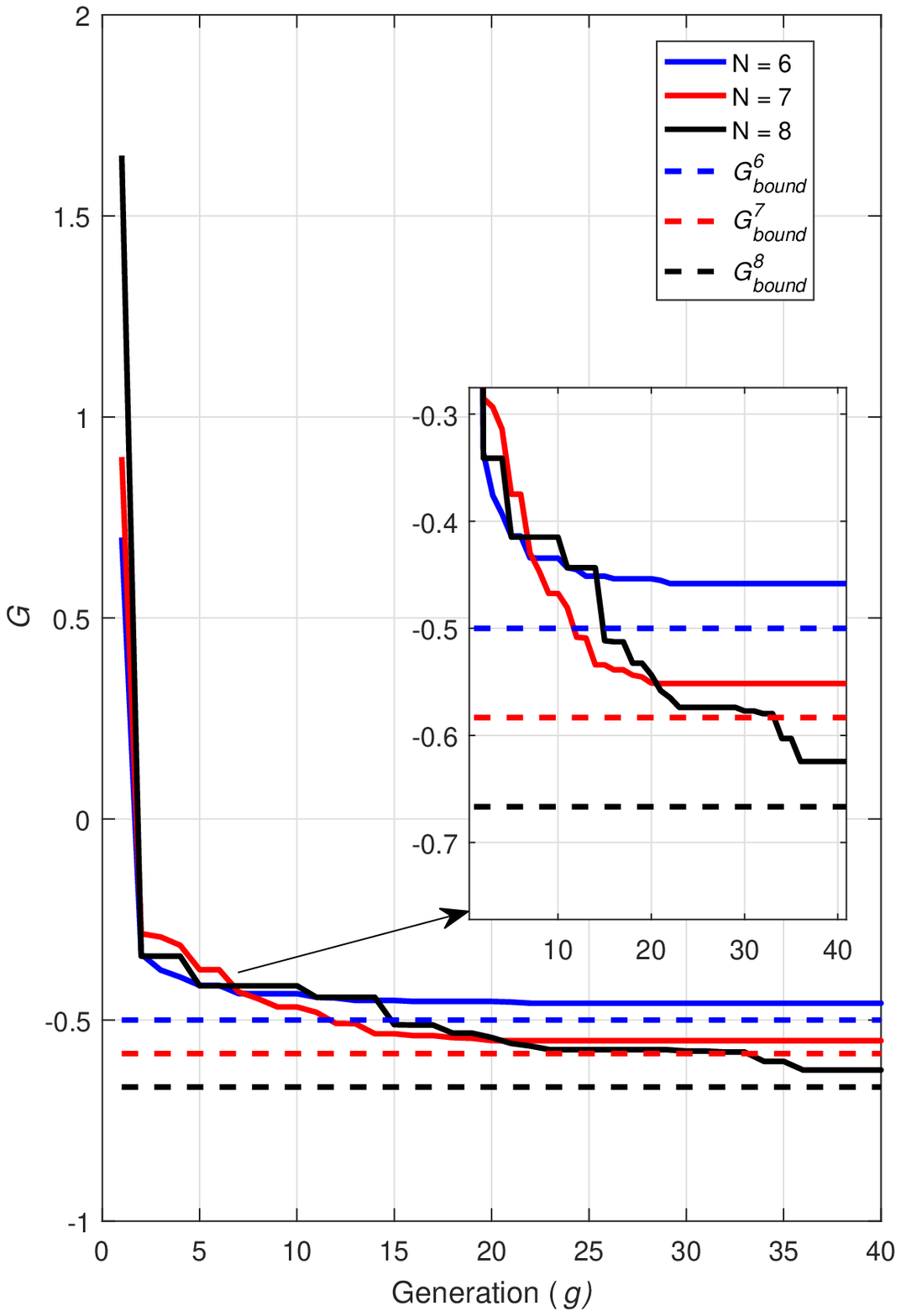}\\
	\quad {\small \hspace{5mm}{(}{\bf a}) Directivity \hspace{16mm} {(}{\bf b}) Objective function $\mathcal{G}$ values}
	\vspace{-1mm}
	\caption{{Directivity and OF values $\mathcal{G}$ through the GA-UPA generations for $N = 6,7$ and $8$. ULA directivity is included in (a);  a lower bound for $\mathcal{G}^N$ following \eqref{eq:Gbound} is defined by dashed lines in (b).}}
	\label{fig:12}
\end{figure}

The values of directivity $\mathcal{D}$ through the generations of the {GA-UPA} is compared in Fig. \ref{fig:12}.a with the regular ULA design; it is noticeable the excellent performance increase in the early generations, being capable of surpassing the conventional ULA design after the first generation.  The convergence of the proposed algorithm is observable with the stationary directivity performance over the generations, which occurs {after $g =  21, 20$ and $35$ generations} for $N = 6, 7$ and $8$ antenna elements, respectively.

The sum of all the {difference-coordinates} points ($x_{mn}$,$y_{mn}$)  {corresponding to the OF  in eq. \eqref{newproblem} typically evolves through generations,} as depicted in  Fig. \ref{fig:12}.b for $N = 6,7$ and $8$; also, lower bound values can be calculated using eq. \eqref{eq:Gbound} and considering $A_n =1,  \, \forall \,  n$.  The bound values $\mathcal{G}_{\text{bound}}^6 = -\frac{1}{2}$, $\mathcal{G}_{\text{bound}}^7 = -\frac{7}{12}$ and $\mathcal{G}_{\text{bound}}^8 = -\frac{2}{3}$ are included for comparison purpose. Hence, after algorithm convergence, the {GA-UPA} performance gap can be calculated as: 
\begin{align}\label{eq:gap}
\Delta \mathcal{G}^N = \mathcal{G}_{\text{bound}}^N - \mathcal{G}_{N}
\end{align}
where $\mathcal{G}_{N}$ is the OF value attained with the GA for $N$ antenna-elements after convergence. Applying \eqref{eq:gap}  on the values found in Fig. \ref{fig:12}.b, the gaps can be established:
$$
\Delta \mathcal{G}^6 = 0.0418; \qquad \Delta \mathcal{G}^7 = 0.0318; \qquad \Delta \mathcal{G}^8 = 0.0423 
$$

The gap indicates the proximity with the best solution possible; hence, for all three simulation scenarios, the GA-UPA method achieves almost the best solution in terms of the performance gap. Moreover, by increasing $N=6$ to 8, it is possible to verify a decrease in $\mathcal{G}$ values for the same number of generations. {Indeed, a} decreasing in $\mathcal{G}$ is responsible for the gain in the directivity,  Fig. \ref{fig:12}.a.

\subsection{{Improved GA-aided UPA Directivity Methods}} \label{sec:GA-aided}
{The GA-aided optimization directivity approach of the previous section demonstrated certain difficulty in finding the optimal solution when the number of antennas increases. Indeed, considering exhaustive combinations in the numerical simulations, the GA-ULA could not outperform the proposed OUPA directivity method,} in a manageable time,  for $N>12$, with the constrain of searching the solution only on the desired plane, which significantly reduces the {GA-ULA complexity.} 

Aiming at using the GA heuristic optimization approach in scenarios with more antennas, we propose changing the GA initialization parameters.  Hence,  to solve the UPA element-antenna position problem in a manageable time, the first approach named GA-{\it marginal} includes an initial population based on a near-local solution.  This approach's main changes are summarized in Table \ref{Table:changes}.

\begin{table}[!htbp]
\centering
\caption{GA-{\it marginal} input parameters}
\begin{tabular}{ll}
\hline
\bf Parameter & \bf Alteration \\
\hline
Initial Population & {\it Near-local} solution \\ 
Mutation ($M_{\text{Uniform}}$) & Uniform Mutation  \\
Population Size ($p_{\text{size}}$) & {$8 N^2$}      \\
Elements placement bounds & $[x_{\max},\, y_{\max}] = [2p_{\max}, \, 2p_{\max}]$  \\
Max. \# Generations ($g_{\text{max}}$) & Unlimited   \\
Stopping Criterion & Outperform OUPA \\
Mutation rate & Determined via simulation \\
Crossover Fraction ($c_{f}$) & Determined via simulation\\
\hline
\end{tabular}\label{Table:changes}
\end{table}

The {\it near-local} solution strategy consists in adding the OUPA solution to the initial population, but with a small perturbation, preventing the GA search from being restricted to the local solutions. Besides,  the {\it mutation} function was changed to be uniform, increasing the diversity to escape local optima.  The {\it population} size was considerably increased to expand the chances of finding the better solution at the cost of increasing algorithm complexity. Moreover, the bounds of element placement were altered considering the OUPA solution, the new values are dependent on $p_{\max}$, which is the max value of the coordinate $x$ or $y$ of the OUPA solution. The {\it stopping} criterion was modified in a way that the $g_\text{max}$ became unlimited, and the new criterion consisted in incrementing the generations until GA-{\it marginal} outperforms the OUPA method. Furthermore, the {\it mutation rate} and the {\it crossover fraction} is now determined by a simulation wherein a combination of both values are evaluated jointly for every 100 generations, and then, the combination that achieves the best value of directivity is selected; the range of both values are set equal and multiple percentages of 10$\%$.  

\vspace{3mm}
\noindent{{\bf Remark 5}: the idea of introducing modifications on the initial population generation in the} GA-aided directivity method aims to facilitate {the guided search to attain incremental} improvements over the OUPA solution, in a manageable time, given the non-convexity nature of the problem. Indeed, the computational difficulty of outperforming our proposed OUPA method in such burden computational conditions is a clear indication of the expedited and advantageous solution given by the proposed OUPA method.
\vspace{3mm}

However, limiting the GA technique to surpass the proposed OUPA marginally will not give us a promising perspective on how much gain one can attain in the UPA directivity context using heuristic evolutionary techniques, and how much the cost to achieve such improvement.  Hence, to address this issue, another GA modification is proposed herein, using almost the same set of parameters described for the GA-{\it marginal}, but changing only the stopping criterion that now consists of achieving hundred consecutive stalled generations, without directivity gain. This method henceforth will be called GA-{\it stall}.

\subsection{Array Directivity Methods: Numerical Comparison} \label{sec:array_Comparison}
Considering a small number of element-antennas yet,  Table \ref{directivity} exposes values of array directivity found by the OUPA method, by GA-aided UPA improved solutions, and the two classical steering vector beamforming, the UCA, and the conventional ULA, as well as an improved directivity method for the omnidirectional UCA proposed in \citep{Huang_2016}.

\begin{table}[!htbp]
\caption{Directivity Methods Comparison for small $N$'s} \label{table2}
\centering
\begin{tabular}{lccc}
\hline \textbf{$N$} & 6 & 8 & 9\\ 
\hline \hline 
\bf OUPA & $11.70$ dBi  &  {$12.91$} dBi  & {$14.12$} dBi \\
\bf GA-{\it marginal}  & {{$11.81$} dBi} &  {{$13.19$} dBi} & {{$14.12$} dBi} \\
{\bf GA-{\it stall}}  & $12.35$ dBi &  $13.49$ dBi & $14.5$ dBi \\
\bf UCA &  $7.96$  dBi &  $8.73$ dBi & $9.17$ dBi \\
{\bf ULA} &  $9.17$   dBi	 &  {$10.38$}  dBi & {$10.88$} dBi \\
{\bf UCA}  \citep{Huang_2016} & n.a & {$12.00$} dBi 	 & n.a.\\
\hline
\end{tabular}\label{directivity}
\end{table} 

The last row in Table \ref{table2} depicts the directivity found in \citep{Huang_2016}, which consists in a technique based on the subspace changes for the omnidirectional UCA directivity. For instance, considering $N=8$,  the maximum directivity value found is close to $\mathcal{D}=12$ dBi. Such value remains remarkably reduced compared with the optimization methods proposed herein, {\it i.e.}, $\sim 1.5$ dBi less compared with the GA{-{\it stall}} approach and $\sim 0.9$ dBi less when compared with our OUPA approach.

\begin{figure}[h]
\vspace{-3mm}
\centering
\includegraphics[clip,width=.4\linewidth]{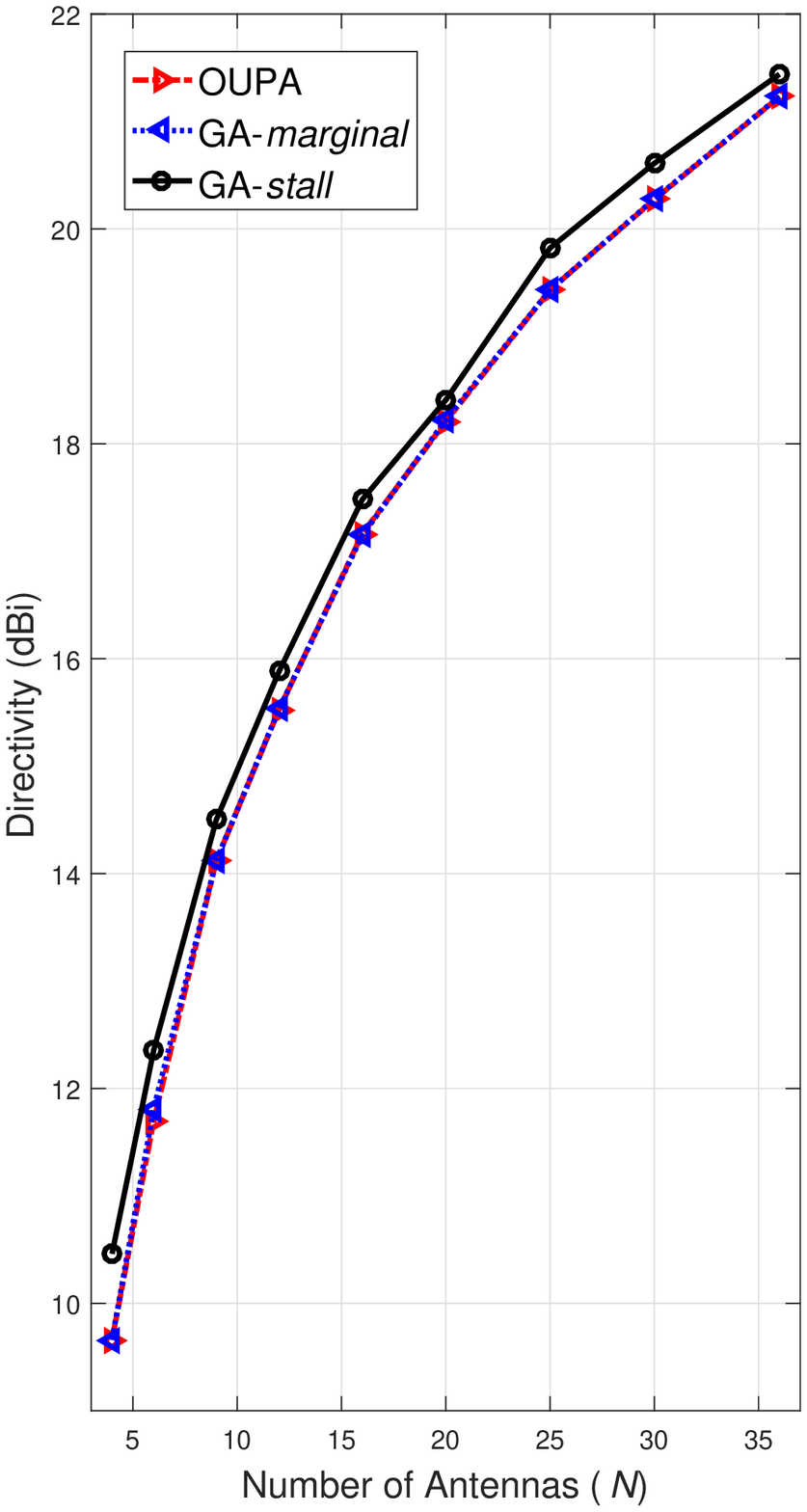}
\includegraphics[clip,width=.4\linewidth]{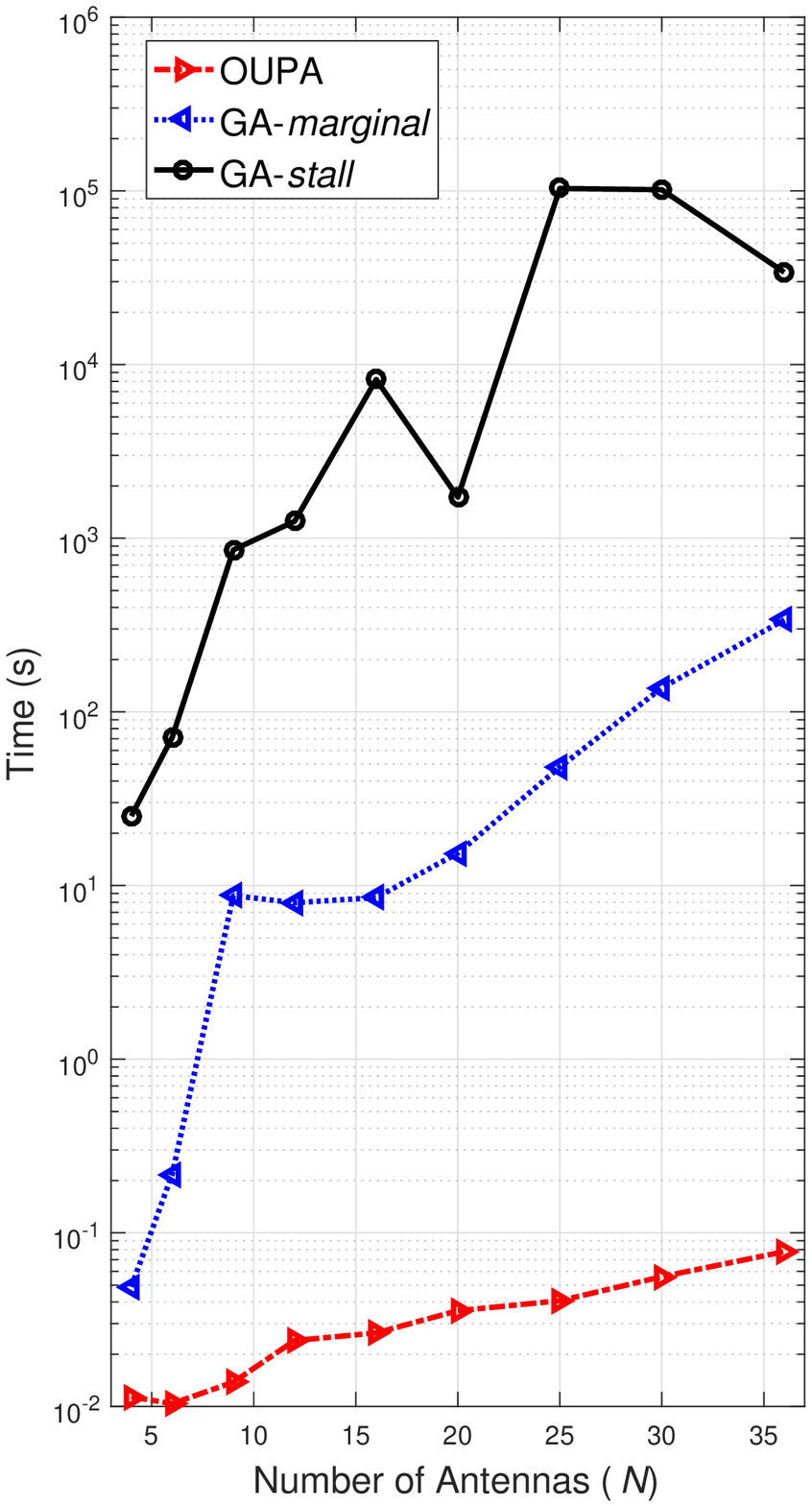} \\
\quad {\small \hspace{1mm} ({\bf a}) Directivity \hspace{25mm} ({\bf b}) {Computational} Time}
\vspace{-1mm}
\caption{Directivity and run-time comparison between OUPA, GA-{\it marginal} and GA with the increasing in the number of antennas $N=N_1\times N_2$.}
\label{fig:aa}
\end{figure}

{For computational complexity analysis, Fig. \ref{fig:aa} exhibits the simulation time {\it vs.}  \# antennas $N\in[4;\, 36]$, considering the OUPA improved GA-marginal and GA-stall methods proposed,  the GA-{\it marginal} and GA-{\it stall}. Aiming to maximize the directivity, the quasi-squared UPA condition in eq. \eqref{n1n2} is applied; hence, possible combinations of $N_1$ and $N_2$ was bounded by $\max(N_1,N_2) = 6,$ implying in the following possible antenna-elements arrangements
\begin{align}N= N_1 \times N_2  \rightarrow 
\begin{array}{ccc}
2 \times 2 & 2 \times 3 & 3 \times 3\\
3 \times 4 & 4 \times 4 & 4 \times 5 \\
5 \times 5 & 5 \times 6 & 6 \times 6 \\
\end{array}, 
\end{align} 
which are identified by the respective markers in Fig. \ref{fig:aa}.}

The OUPA technique proposed herein, even with far less complexity, can follow the performance of both GA's techniques very closely.  As expected, the GA-{\it marginal} was able to surpass the OUPA method marginally, proving its sub-optimality. However, the time spent by both GA heuristic methods was far superior, configuring a far inferior {performance}-complexity tradeoffs. The GA-{\it stall aided UPA directivity optimization method} was able to achieve a considerable and consistent improvement when compared with the OUPA, the amount of time consumed to find such a solution is extensively large. 

\vspace{3mm}
\noindent{{\bf Remark 6}:} One can conjecture that {among the three proposed directivity methods,} the improvement beyond the OUPA solution provided by the GA-{\it stall} aided UPA algorithm comes with a much high computational cost. Moreover, both GA-{\it stall} and GA-{\it marginal} optimization schemes have resulted in less effectiveness than the proposed OUPA method in terms of performance-complexity tradeoff.
\vspace{3mm}

Finally,  Table \ref{Table:parameters} shows the {\it crossover fraction} ($c_f$) and the {\it mutation rate} ($m_r$) values as a function of dimensionality problem ($N$), required in the GA-aided methods that maximize the directivity in the simulation of Fig. \ref{fig:aa}.
\begin{table}[!htbp]
\centering
\caption{{Input parameters configuration for both GA-{\it marginal} and GA-{\it stall} algorithms}}
	\begin{tabular}{lcccccccccc}
		\hline
		\bf N & \bf 4 & \bf 6 & \bf 9 & \bf 12 & \bf 16 & \bf 20 & \bf 25 & \bf 30 & \bf 36\\
		\hline
		\vspace{-0.1cm} & \vspace{-0.1cm} \\
		{$c_f$ ($\%$) } & 80  & 40  & 10  & 90  & 40  & 80  & 80  & 60 & 80 \\	
		{$m_r$ ($\%$)} & 100  & 20  & 10  & 10  & 10  & 70  & 10  & 10 & 20  \\	
		\hline
	\end{tabular}\label{Table:parameters}
\end{table}
For most configurations, the crossover fraction percentage has high values, while the mutation rate presents a low percentage. Such a combination lead to a population that has several changes to escape local optima between the generations, nonetheless, the value of $m_r$ remaining low can be interpreted as a mean of guaranteeing an improvement around the solution. Notice that in the case $N=20$, both GA-{\it stall} directivity optimization algorithms selected a high  $m_r$ percentual value; as a result, the simulation time resulted significantly smaller; however, the performance was incremental better when compared to the OUPA and GA-{\it marginal}. The discrepant values found for $N=4$ antennas can be explained by the simplicity of the geometric configuration, therefore, given a population, a volatile evolution did not prevent the method from finding a remarkable solution.

\section{Conclusions {and Final Remarks}} \label{sec:concl}
This work proposes a new approach to maximize the directivity of an omnidirectional volumetric antenna array.  This technique assumes a uniform planar array (UPA) confined on a specific plane with a minimum distance between the antenna elements optimized by the successive evaluation and validation (SEV) procedure, namely {\it optimal uniform planar array} (OUPA).  Moreover,  evolutionary heuristic GA optimization is employed to validate the proposed methodology;  hence, for a  large number of antennas, two modifications on the initial parameters are suggested, denoted GA-{\it marginal} and GA-{\it stall}, both made the GA a promising optimization tool to solve UPA directivity problem in such large scale antenna scenario.

An exciting finding is that a plane space constrains the element positioning solutions as a function of the desired elevation ($\theta_0$) and azimuth ($\phi_0$) angles.  A new method based on the UPA structure is introduced to address this constraint to solve the directivity optimization problem with low-complexity effort. Then an evolutionary heuristic technique is selected to implement the directivity optimization analysis devised herein.  Indeed, the genetic algorithm (GA) was selected as an expedited optimization tool. 

The proposed methodology is focused on finding the optimal position of each antenna element that maximizes the antenna array directivity.  The solution was confined to a specific plane, where the antenna elements position on this plane resembling a sequence of equilateral triangles; hence, the idea of finding the optimal UPA in such a plane is introduced. The OUPA method exploits the OF features, while in the evolutionary heuristic optimization approach, the directivity based on antenna-elements position has been formulated and solved for a different number of antenna-elements in the range $N \in \{4;\, 36\}$. Indeed, numerical results deploying the proposed GA-{\it marginal} and GA-{\it stall}  heuristic evolutionary techniques against the also proposed OUPA method for a different number of antennas has demonstrated higher directivity gains when compared with the well-known regular UCA and ULA arrangements, as well as when compared with recent literature design for the specific case of $N=8$ antennas. 

The computational complexity {reveals} a superior performance of the OUPA method, implying in an excellent performance-complexity tradeoff.  Moreover, the OUPA design method has achieved impressive performance in massive MIMO scenarios ($N\geq 50$), achieving a directivity of $30$ dBi in an extensive quasi-squared planar configuration, {\it i.e.}, $N = 15 \times 16$ antennas.

\section*{Acknowledgement}
This work was partly supported by The National Council for Scientific and Technological Development (CNPq) of Brazil under Grants 310681/2019-7, partly by the CAPES- Brazil - Finance Code 001,  and the Londrina State University - Parana State Government (UEL).

\end{document}